\documentclass[11pt]{article}
\usepackage[iso-8859-7]{inputenc}
\usepackage{epsfig}
\usepackage{graphicx}
 \usepackage{cite}
 \usepackage{float}
 \usepackage{amsmath} 
 \usepackage{amssymb}   
  \usepackage{slashed}
 \usepackage{amsthm} 
 \usepackage{geometry}
 \usepackage{graphicx} 
 \usepackage{multirow} 
\usepackage{amsmath}
\usepackage{braket}
\usepackage{mathtools}
\usepackage[compat=1.1.0]{tikz-feynman}
\usepackage{subcaption}
\usepackage{array}

 \usepackage{scalerel}
\usepackage{tikz}
\usetikzlibrary{svg.path}

\definecolor{orcidlogocol}{HTML}{A6CE39}
\tikzset{
  orcidlogo/.pic={
    \fill[orcidlogocol] svg{M256,128c0,70.7-57.3,128-128,128C57.3,256,0,198.7,0,128C0,57.3,57.3,0,128,0C198.7,0,256,57.3,256,128z};
    \fill[white] svg{M86.3,186.2H70.9V79.1h15.4v48.4V186.2z}
                 svg{M108.9,79.1h41.6c39.6,0,57,28.3,57,53.6c0,27.5-21.5,53.6-56.8,53.6h-41.8V79.1z M124.3,172.4h24.5c34.9,0,42.9-26.5,42.9-39.7c0-21.5-13.7-39.7-43.7-39.7h-23.7V172.4z}
                 svg{M88.7,56.8c0,5.5-4.5,10.1-10.1,10.1c-5.6,0-10.1-4.6-10.1-10.1c0-5.6,4.5-10.1,10.1-10.1C84.2,46.7,88.7,51.3,88.7,56.8z};
  }
}

\newcommand\orcidicon[1]{\href{https://orcid.org/#1}{\mbox{\scalerel*{
\begin{tikzpicture}[yscale=-1,transform shape]
\pic{orcidlogo};
\end{tikzpicture}
}{|}}}} 

 \usepackage{hyperref}

\newcommand{\newc}{\newcommand}
\newc{\ra}{\rightarrow}
\newc{\lra}{\leftrightarrow}
\newc{\ov}{\overline}
\newc{\pa}{\partial}
\newc{\be}{\begin{equation}}
\newc{\ee}{\end{equation}}
\newc{\ba}{\begin{eqnarray}}
\newc{\ea}{\end{eqnarray}}

\newc{\D}{\Delta}
\newc{\vt}\vartheta
\newc{\nn}{\nonumber}

\begin{document}
	\thispagestyle{empty}
\begin{titlepage}
	\vspace*{0.7cm}
	\begin{center}
			{\Large {\bf Sterile neutrinos, $0\nu\beta\beta$ decay and the W-boson mass anomaly in a Flipped $SU(5)$ from F-theory}}
		\\[12mm]
		Vasileios Basiouris${^{\diamond}}$~\orcidicon{0000-0002-9266-7851}~\footnote{E-mail: \texttt{v.basiouris@uoi.gr}},
		George K. Leontaris${^{\diamond,\star}}$~\orcidicon{0000-0002-0653-5271}~\footnote{E-mail: \texttt{leonta@uoi.gr}}
	\end{center}
\vspace*{0.20cm}
\centerline{\it
${^{\diamond}}$~Physics Department, University of Ioannina}
\centerline{\it 45110, Ioannina, 	Greece}
\vspace*{0.20cm}
\centerline{\it
${^{\star}}$~Physics Department, CERN}
\centerline{\it 1211, Geneva, 23, 	Switzerland}
\vspace*{1.20cm}

	\begin{abstract}
			
	We investigate the low energy properties of an effective   local model   with flipped  $SU(5)\times U(1)_{\chi}$ gauge group, constructed within  the framework of F-theory.  Its  origin is traced back to the $SO(10)$  symmetry 
	-associated with a geometric singularity of the compactification manifold- 
	broken by  an internal flux  which is turned on along the seven-brane in the $U(1)_{\chi}$ direction. Topological properties and the choice of flux parameters determine the massless  spectrum of the model to be that of the minimal flipped $SU(5)$ supplemented with an extra right-handed electron-type state and its complex conjugate, $E^c+\bar E^c$, as well as   neutral singlet fields.
	The  subsequent symmetry breaking to the $SU(3)\times SU(2)\times U(1)_Y$ gauge group occurs with a Higgs  pair in  $10+\overline{10}$ representations of $SU(5)$. 
	Next we proceed to the phenomenological analysis of the resulting effective model 
	and the salient outcomes are: The $E^c+\bar E^c$ pair  acquires a mass of few TeV and as such   could solve the $g_{\mu}-2$ discrepancy.    Neutrino couplings to extra neutral singlets lead to an inverse seesaw mechanism where an extra light state could be a suitable dark matter candidate.  
	The  predictions of the model for the  ${0\nu}\beta\beta$ decay  rate could be tested in near future experiments.  There are non-unitarity deviations from the lepton mixing matrix ($U_{PMNS}$), which could in principle explain the new precision measurement of the W-boson mass  recently  reported  by the CDF II collaboration.

	\end{abstract}
	
 \end{titlepage}

\section{Introduction }

String model building is significant for it turns the superstring data  to low energy predictions which can be  tested experimentally.
Over the last few decades, many models have been engineered in the context of heterotic  and type IIA string theories,    as well as 
IIB  and its geometric analogue, the F-theory.  
The case of F-theory~\cite{Vafa:1996xn}  in particular, establishes  a  robust framework  for model building  by providing the necessary tools for a convenient  
implementation of the string rules and principles in order to construct a  viable effective field theory model with predictive power.
Since many decades, it is widely accepted that the low energy  (Standard Model) matter content is embedded  only in a few representations 
of Grand Unified Theories (GUTs), such as $SU(5), SO(10)$ and the like. In F-theory such GUTs are considered in the context of the exceptional 
gauge group $E_8$ associated with the highest  geometric  singularity of the  internal (compactification) manifold.  As a consequence, in F-theory
 compactifications the geometrical properties of these singularities  encode all the information regarding the  properties of the effective theory. 
 Thus, the observable gauge group is expected to be a subgroup of $E_8$, arising on the worldvolume of a seven-brane wrapping a four-manifold 
 `surface' of the internal six-dimensional space. The geometrical picture  is complemented by the presence of certain seven-brane configurations
  intersecting over the wrapped surface representing the specific GUT. The  gauge sector of the theory is localized on the world volume of the GUT 
  seven-brane whereas matter fields reside on Riemann surfaces (called `matter curves' hereafter) formed in the intersections of other seven-branes with the GUT `surface'. 
  
  \noindent 
   A well   known issue in string derived models, however, is the mismatch between the string scale $M_{string}$ at which  the  above picture is formulated, and 
  the gauge coupling   unification scale $M_{GUT}$ which is found to be two orders of magnitude smaller than  $M_{string}$. Interestingly, in F-theory 
a decoupling of these two scales can be naturally  achieved by  requiring  the spacetime filling seven-brane to  wrap a del Pezzo surface. Therefore, 
in such a scenario gravity is decoupled at a scale higher than the  GUT   point where the three Standard Model gauge couplings unify. 

In many cases of string constructions (including the case of compactifications on del Pezzo surfaces) the available representations for the Higgs sector are
 restricted only to the fundamental and the spinorial ones,  whereas  GUTs such as the standard  $SU(5)$ and SO(10) require the adjoint or higher ones 
 to break the corresponding symmetry. Remarkably, in  F-theory, given a GUT with gauge group $G_S$,  the  symmetry breaking can occur by developing a flux along a 
 $U(1)$ factor inside  $G_S$. The symmetry breaking of $G_S=SU(5)$ for example,  is achieved by fluxes turned on along the hypercharge $U(1)_Y$ factor~\footnote{A natural question arises whether the corresponding gauge boson remains massless. According to~\cite{Donagi:2008ca,Beasley:2008dc,Donagi:2008kj,Beasley:2008kw} a necessary and sufficient  topological condition for the $U(1)_Y$ gauge boson to remain massless is a non-trivial cohomology class of the flux on the seven-brane  while it represents a trivial class in the base of the F-theory
	compactification.}. 
Moreover, the restriction of fluxes along the matter curves split the $SU(5)$ representations and determine the multiplicity of the matter content in terms of a few integers associated with those fluxes.
For higher gauge groups, however, it  is expected that fluxes must be turned on along more than one $U(1)$ factors to fully break the GUT symmetry down to the Standard Model one. In $SO(10)\supset SU(5)\times U(1)_{\chi}$  for example,  a flux  can be turned on
along the $U(1)_{\chi}$ factor breaking it the to $SU(5)$ symmetry
which subsequently is reduced to the Standard Model one in the way described above. 
However, because of possible restrictions on the integer parameters associated with the fluxes, the combined effects of successive 
flux-induced breaking mechanisms might lead to unsought particle content.  Moreover, topological constraints may not support integer fluxes
for successive flux symmetry breaking along multiple $U(1)$ factors.   
In that respect, it would be desirable to investigate whether the  GUT groups beyond the minimal $SU(5)$ could be reduced to the standard model symmetry by combining both flux and Higgs symmetry breaking mechanisms.  In the present work, we follow  this approach in the case of  $SO(10)$ gauge group. The appealing features of this model are well known. Among others, it is the only GUT where all the matter  fields (including the right-handed neutrino) of each generation,  are accommodated in a single $SO(10)$ representation, namely the spinorial $\underline{16}$. In the standard GUT approach,
  where all representations are available, the most familiar
 Higgs symmetry breaking patterns of $SO(10)$ are through the intermediate
symmetries of $SU(5)\times U(1)_{\chi}$ and the left-right symmetric $SU(4)\times SU(2)\times SU(2)$ Pati-Salam symmetry. Interestingly, both of them  do not require large Higgs representations to break down to the standard model group. 
The first case, is identified with the well known 
flipped  $SU(5)\times U(1)_{\chi}$ model which requires only
the $10+\overline{10}$ Higgs fields for its breaking\cite{Barr:1981qv,Antoniadis:1987dx}. Also, the breaking of the Pati-Salam symmetry\cite{Pati:1974yy}  can be realized
 by the vector-like Higgs pair of fields which  transform as
$(4, 1, 2)+ (\bar 4, 1, 2)$~\cite{Antoniadis:1988cm}.

In the present study, we will investigate the low energy implications of the effective model derived under the  first  symmetry breaking chain discussed above. Thus, we will consider  its embedding in the highest ($E_8$)  geometric singularity, with an $SO(10)$ divisor, so that 
\be 
SO(10)\supset  SU(5)\times U(1)_{\chi} \supset SU(3)\times SU(2)\times U(1)_Y\label{SO10chain}
\ee 

According to the previous discussion, the first stage of symmetry breaking will occur with the flux mechanism by turning on fluxes along the $U(1)_{\chi}$. Provided that we define the hypercharge generator as the appropriate linear combination of  $U(1)_{\chi}$ and the abelian factor inside $SU(5)$, the resulting theory is exactly  the flipped $SU(5)$ model. The fermion particle content in 
particular is accommodated in the $10+\bar 5+1$ descending  from
the $\underline{16}$ of $SO(10)$.  Further, there are  Higgs fields in $10+\overline{10}$ of $SU(5)$ descending from the $\underline{16}+\underline{\overline{16}}$ and there are  $5+\overline{5}$ coming from the $\underline{10}$ of $SO(10)$. Then, the standard model 
gauge symmetry is obtained when vacuum expectation values (VEVs) are developed along the
pair $10+\overline{10}$ whilst $5+\overline{5}$  provide the Higgs doublets.

We note in passing that another important aspect of the flipped model in F-theory, is
that we could equivalently trace its origin  through the $SU(5)$ symmetry and a Mordell Weil $U(1)$ symmetry (for reviews see~\cite{Weigand:2018rez,Cvetic:2018bni}).
 This would bring additional discrete symmetries, some of them being  of the type $Z_m\times Z_n$,
 which could be useful for yet unconstrained Yukawa Lagrangian terms. We leave this investigation for
 a future work and here we only focus on the derivation of flipped through its embedding in $E_8$
 leading to the symmetry breaking chain~(\ref{SO10chain}). 
 
  Once we have derived the final gauge symmetry by combining flux and Higgs mechanisms,  we focus on  the zero-mode spectrum of the model, the Yukawa potential and its basic properties.   Next we explore the implications in a wide range of processes being of current interest. Thus, among others, we analyze the predictions in  neutrino physics, proton decay, leptogenesis and double beta decay. We further  discuss  a potential interpretation of the recently detected anomaly on the W-boson mass as observed by the CDF experiment. 
  
 The layout of the paper is as follows. In order for this work to be self-contained, in section 2 we present a short introduction on the field theory flipped $SU(5)$. In section 3 we present a semi-local version of the model from F-theory. We introduce a $U(1)$ flux to break   $SO(10)$ down to the flipped $SU(5)\times U(1)$ symmetry whilst, subsequently, we implement the Higgs mechanism to reduce the gauge symmetry down to the Standard Model one.
 In section 4 we present the superpotential and its basic low energy properties. In section 5 we derive some bounds on the parameter space from non-observation of proton decay. In section 6 we investigate the form of  the neutrino mass matrix and show that it acquires a type II form   due to mixing  of ordinary neutrinos with  additional (inert) singlet fields.
 We discuss various limiting cases and determine the
 conditions so that a few keV neutral state appears to play the role of dark matter. Moreover, the implications on the leptogenesis scenario are investigated. Section 7 is devoted in a detailed consideration of double beta decay within the flipped $SU(5)$ context. 
 In section 8 we discuss the $g_{\mu}-2$ anomaly and in section 9 we present a possible interpretation of the W-mass new measurement  recently determined by CDF collaboration. In section 10, we discuss the renormalization group evolution and we also address the effect of the vector-like family in the Yukawa couplings. We present our conclusions in section 11 and include some computational details in the Appendix.



\section{$SU(5)\times U(1)$ basics }

 We would like to investigate the flipped $SU(5)\times U(1)$ model in a generic F-theory framework. 
 Within the proposed framework
we  implement the spectral cover  approach and turn on   fluxes along $U(1)$'s to determine 
the geometric properties of the matter curves and  the massless spectrum  residing  on them. At this  
stage we end up with the  flipped $SU(5)$ which we  envisage it contains the three generations of the chiral matter fields,
and the necessary Higgs representations to break the symmetry.

Before we attempt to derive this model from F-theory, we give a brief  account of the field theory version. 
The chiral matter fields of each  family  constitute a complete $\underline{16}$ spinorial representation  of $SO(10)$   
which  admits the  $SU(5)\times U(1)_{\chi}$ decomposition  
\be 
\underline{16}=10_{-1}+\bar 5_3+1_{-5}~.\label{16dec}
\ee
Denoting with $x$ the `charge' under $U(1)_{\chi}$ and $y$ 
under the $U(1)$ of the familiar Standard Model symmetry group, 
the hypercharge definition for flipped $SU(5)$  is 
$Y=\frac 15\left(x+\frac 16 y\right)$. 
This  implies  the following  embedding of the Standard Model representations
\ba
10_{-1}&\Rightarrow&F_i\;=\;(Q_i,d^c_i,\nu^c_i)\\
\bar 5_{+3}&\Rightarrow&\bar f_i\;=\;(u^c_i,\ell_i)\\
1_{-5}&\Rightarrow&\ell^c_i\;=\;e^c_i~.
\ea
As already pointed out, the spontaneous symmetry breaking of 
 the flipped $SU(5)$ symmetry  occurs with a pair of  Higgs fields accommodated in 
\ba
H\equiv 10_{-1}\;=\;(Q_H,d_H^c,\nu_H^c)&,&\bar H\equiv   \ov{10}_{+1}\;=\;(\bar Q_H,\bar d_H^c,\bar \nu_H^c)~.\label{10H}
\ea 
The MSSM Higgs doublets are found in the fiveplets descending
from the $\underline{10}$ of $SO(10)$ 
\ba 
h\equiv 5_{+2}\;=\;(D_h,h_d)&,&\bar h\equiv \bar 5_{-2}\;=\;(\bar D_h,h_u)~.
\label{10toh5h5bar}
\ea
A remarkable fact in the case of the flipped model is that
the $U(1)_{\chi}$ charge assignment distinguishes the  Higgs  $\bar 5_{-2}$ fields from  matter anti-fiveplets $\bar 5_3$. In particular, the former  contain down-quark type triplets $\bar D_h$ while the latter accommodate the $u^c$ quarks.

The fermion masses arise from the following $SU(5)\times U(1)_{\chi}$ invariant couplings
\ba  
{\cal W}&\supset&\lambda_d\,  10_{-1}\cdot 10_{-1}\cdot 5^h_{2}+\lambda_u\, 10_{-1}\cdot\bar 5_{3}\cdot\bar 5^{\bar h}_{-2}+\lambda_{\ell}\,1_{-5}\cdot\bar 5_{3}\cdot 5^h_2
\\
&\supset&\lambda_d\, Q\,d^c\,h_d+\lambda_u\, (Q\,u^c\,h_u+\ell\nu^c\,h_u)+\lambda_{\ell}\, e^c\,\ell\,h_d~.
\ea 
It should be observed that the flipped  model at the GUT scale predicts  that up-quark and neutrino Dirac mass matrices
are linked to each other and in particular, $m_t=m_{\nu_{\tau}}$.  
However, in stark contrast to the standard $SU(5)$ model, down quarks and lepton mass matrices are 
unrelated, since in the flipped model  they originate from different Yukawa
couplings.

Proceeding with the Higgs sector, as  $H_,\,\bar H$ acquire large VEVs of the order $M_{GUT}$,
they break  $SU(5)\times U(1)_\chi$ down to Standard Model gauge group and 
at the same time they provide heavy masses to the color triplets.
Indeed, the following  mass terms are obtained
\ba
HHh+\bar H\bar H\bar h\;
\Rightarrow \;
\langle \nu^c_H\rangle d^c_{H}D+\langle \overline{\nu_H^c}\rangle \bar d^c_{H}\bar D~.
\ea

 Moreover, a higher order term providing right-handed neutrinos with Majorana masses is of the form
\be
\label{Maj}
\begin{split}
{\cal W}_{\nu^c}&=\frac{1}{M_S}\overline{10}_{\bar H}\overline{10}_{\bar H}\,10_{-1}\,10_{-1}\\
&=  \frac{1}{M_S}\overline{H}\overline{H} F_i F_j\;
\Rightarrow  \; \frac{1}{M_S}\langle \overline{\nu_H^c}\rangle^2\nu_i^c \nu_j^c~.
\end{split}
\ee
It should be noted that possible couplings with additional neutral singlets $\nu_{s}$  may extend the seesaw mechanism to type II. As we
will see, this is exactly the case of the F-theory version.

\section{Flipped from F theory }

In the context of local  F-theory constructions we may assume
an $E_8$ point of enhancement  where the  flipped $SU(5)$ 
emerges  through the following 
symmetry reduction
\be 
E_8 \supset SO(10)\times SU(4)_\perp\supset [SU(5)\times U(1)]\times SU(4)_{\perp}~,
\label{E8to514}
\ee 
where $SU(4)_{\perp}$ incorporates the symmetries of  the spectral cover.
Matter fields are accommodated in irreducible representations 
emerging from the decomposition of the $E_8$ adjoint under $SO(10)\times SU(4)$ 
\ba
248&\ra& (45,1)+(1,15)+(10,6)+(16,4)+(\overline{16},\overline{4})~,
\ea
followed by the familiar reduction of $SO(10)$ representations given in (\ref{16dec}) and (\ref{10toh5h5bar}),
according to the second stage of breaking $SO(10)\to SU(5)\times U(1)$ as shown in~(\ref{E8to514}).
The following  invariant trilinear couplings provide with 
masses up and down quarks, charged leptons and neutrinos
\ba
{\cal W}_{down}&\in&(10,4)_{-1}\cdot (10,4)_{-1}\cdot (5, 6)_{2}\label{dsec}\\
{\cal W}_{up/\nu}&\in&(10,4)_{-1}\cdot (\bar 5,4)_{3}\cdot (\bar 5, \bar 6)_{-2}\label{usec}\\
{\cal W}_{\ell}&\in& (1,4)_{-5}\cdot (\bar 5,4)_3\cdot (5,6)_2~.\label{lsec}
\ea
As opposed to the plain field theory model,  the corresponding
trilinear couplings transform non-trivially under the spectral 
cover $SU(4)_{\perp}$  group. However, the matter fields  reside 
 on 7-branes whose positions are located at the singularities of the fibration. In the geometric language of F-theory constructions,  the 
 matter fields of the effective model  are found on the matter
 curves  where the gauge $SU(5)\times U(1)$ symmetry is appropriately  enhanced. Moreover, 
 their corresponding trilinear Yukawa couplings are formed at the intersections of three matter curves where the symmetry is further enhanced.  In the spectral cover picture the   symmetry enhancement of each representation can be described by the  appropriate element of the $SU(4)_{\perp}$ Cartan subalgebra which is parametrized by four weights $t_i$ satisfying   $\sum_{i=1}^4 t_i=0$. The latter are associated with the roots
 of a fourth degree polynomial related to the $SU(4)_{\perp}$ spectral cover. The coefficients of this polynomial equation
  convey information related to the geometric properties of the fibred  manifold to the effective theory. 
 Usually, there are non-trivial monodromies~\cite{Heckman:2009mn} identifying roots of
 the fourth degree  polynomial equation 
 associated with $SU(4)_{\perp}$. In the 
 present case the identification of matter curves occurs through  a discrete
 group  which is a subgroup of the maximal 
 discrete (Weyl) group $S_4$ of  $SU(4)_{\perp}$.

To proceed, first we identify the weights of matter field 
representations. At the $SO(10)$ level, the $\underline{16}$
 transforms in $4\in SU(4)_{\perp}$ and  $\underline{10}$ 
in $6\in SU(4)_{\perp}$ so we make the following identifications
\be 
\label{5and10}
\begin{split}
 (\underline{16}, 4)&\to   \underline{16}_{t_i},\; i=1,2,3,4\\  (\underline{10}, 6)&\to  \underline{10}_{t_i+t_j} \; i,j=1,2,3,4~.
\end{split}
\ee 
In principle, there are four matter curves to accommodate $ \underline{16}+\overline{\underline{16}}$ representations and six for the $ \underline{10}$'s of $SO(10)$.
We will focus on the phenomenologically viable case  of  the minimal $Z_2$  monodromy. This choice implies rank-one  mass matrices  where only the third family of quarks are present at tree-level ensuring a heavy  top-quark mass in accordance with the experiments. Thus,  implementing the $Z_2$ monodromy by imposing the
 identification of  the two weights $t_1\leftrightarrow t_2$, the  matter
curves of (\ref{5and10})  reduce to 
\be 
\label{5and10Z2}
\begin{split}
 \underline{16}_{t_i}&\to    \underline{16}_{t_1},\, \underline{16}_{t_3},\, \underline{16}_{t_4}
\\
 \underline{10}_{t_i+t_j} &\to   \underline{10}_{2 t_1},\, \underline{10}_{t_1+t_3},\, \underline{10}_{t_1+t_4}\, \underline{10}_{t_3+t_4}~.
\end{split}
\ee

\subsection{${\cal Z}_2$ monodromy}

Information regarding the geometric properties of the
matter curves and the representations accommodated on them
can be extracted from the polynomial equation  
for the $SU(4)$ spectral cover. This equation is 
\be 
\sum_{k=0}^4b_ks^{4-k}=b_0 s^4+b_1 s^3+b_2s^2+b_1 s^3+b_4=0~.\label{SC4}
\ee
The coefficients $b_k$ are sections of $[b_k]=\eta -k c_1$
while we have defined $\eta =5c_1-t$ with $c_1$ ($-t$) being the $1^{st}$ Chern class of the tangent (normal) bundle to the GUT `surface'.
 Under the assumed $Z_2$ monodromy  the   spectral cover equation is
 factorized as follows
\be
\begin{split}
	{\cal C}_4&=(a_1+a_2s+a_3s^2)(a_4+a_5s)(a_6+a_7s)
\\
&=a_1a_4a_6+(a_1a_5a_6+a_2a_4a_6+a_1a_4a_7)s
\\
&+(a_1a_5a_7+a_2a_5a_6+a_3a_4a_6)s^2+(a_3a_5a_6+a_2a_5a_7)s^3+a_3a_5a_7s^4~.
\label{z2case}
\end{split}
\ee
Comparing this to (\ref{SC4}) we extract equations of the form
$b_k =b_k(a_i)$
\be 
\label{asfrombs}
\begin{split}
b_4&=a_1a_4a_6\\
b_3&=a_1a_5a_6+a_2a_4a_6+a_1a_4a_7
\\
b_2&=a_1a_5a_7+a_2a_5a_6+a_3a_4a_6
\\
b_1&=a_3 a_5 a_6+a_3 a_4 a_7+a_2 a_5 a_7   
\\
b_0&=a_3a_5a_7,
\end{split}
\ee 
 and use them to derive the relations for
the homologies $[a_i]$ of the coefficients $a_i$. There are five equations relating $b_k$'s with products of $a_i$ coefficients and all five of them can be cast in the form
\ba
\eta-k\,c_1&=&[a_l]+[a_m]+[a_{n}],\;\; {\rm where} \;\; k+l+m+n=15~,
\label{homolequs}
\ea
with $k=0,1,2,3,4$ and  $l,m,n$ take the values $1,2,\dots, 7$.
For example, the  term $a_3a_4 a_6s^2$ in (\ref{z2case}) gives
$[a_3]+[a_4]+[a_{6}]+2[s]=(\eta -2 c_1)-2c_1= c_1-t$ and 
analogously for the other terms. 
 The system (\ref{homolequs}) consists of  five linear equations involving products of the coefficients $a_i$ with yet unspecified homologies $[a_i]$  which  must be determined  
   in terms of the known $[b_k]$.  Since there are  five linear equations with seven unknowns we 
   can  express $[a_i]$  in terms of two arbitrary
parameters defined as follows:
$$\chi_5=[a_5], \chi_7=[a_7],\; \chi =\chi_5+\chi_7~. $$
Then, we find that 
\[ [\alpha_i]=\eta -(3-i)c_1-\chi,\,i=1,2,3\;;\; [a_5]=[a_4]+c_1=\chi-\chi_7\;;\; [a_7]=[a_6]+c_1=\chi_7~.
\]
Note that because of the vanishing of the coefficient $b_1=0$, we also need to solve the constraint $b_1(a_i)=0$. It can be readily seen that a possible solution is achieved by defining a new section $k$ with 
$[\kappa]=\eta-2\chi$ such that
\be
\label{Ansatz}
\begin{split}
	a_3&=\kappa a_5 a_7,\;\; a_2=-\kappa (a_5a_6+a_4a_7)~.
\end{split}
\ee
 Using the above 
topological data we can now specify the flux restrictions on the matter curves  and 
 determine  the multiplicities of the zero mode spectrum and other properties of the effective
field theory model.

\noindent 

From the first of equations~(\ref{asfrombs}), the  condition $b_4=0$  becomes $a_1a_4a_6=0$,  which defines  three  $ \underline{16}$'s localized at $$a_1=0,\; a_4=0,\; a_6=0~.$$
\\
 Similarly, the equation
$b_3^2(a_i)=0$ determines the topological  properties and the multiplicity  of $ \underline{10}$'s.
Substituting~(\ref{Ansatz}) into $b_3$, we obtain
\[\left(a_5 a_6+a_4 a_7\right) \left(a_1- \kappa a_4 a_6\right)=0~.
\]
Knowing the homologies of the individual $a_i$'s we can compute
those of the various matter curves. The  results are shown  in the fifth column of Table~\ref{T_ALL} where  for convenience homologies are parametrized with respect to  
the free parameters  $\chi_5,\chi_7, \chi=\chi_5+\chi_7$.
\begin{table}
	\begin{center}
		\begin{tabular}{|l|l|l|l|l|}
			Matter&$t_i$ charges& Section& Homology &$U(1)_{\chi}$\\
			\hline
			$\underline{16}$&$t_{1}$&$a_1$&$\eta-2c_1-\chi$&$M-P$
			\\
			$\underline{16}$&$t_3$&$a_4$&$-c_1+\chi_5$&$P_5$
			\\
			$\underline{16}$&$t_4$&$a_6$&$-c_1+\chi_7$&$P_7$
			\\
			$ \underline{10}$&$t_{1}+t_3$&$a_1-\kappa a_4a_6$&$\eta-2c_1-\chi$&$M-P$
			\\
			$\underline{10}$&$t_{1}+t_4$&$a_1-\kappa a_4a_6$&$\eta-2c_1-\chi$&$M-P$
			\\
			$\underline{10}$&$2t_{1}$&$a_5a_6+a_4a_7$&$-c_1+\chi$&$P$
			\\
			$\underline{10}$&$t_{3}+t_4$&$a_5a_6+a_4a_7$&$-c_1+\chi$&$P$
			\\
			\hline
		\end{tabular}
	\end{center}
	\caption{ Properties of $SO(10)$ representations in the ${\cal Z}_2$ monodromy.}
	\label{T_ALL}
\end{table}

As already noted,  the $ SO(10)\to SU(5)\times U(1)_{\chi}$  breaking is
achieved by turning on a $U(1)_{\chi}$ flux.  At the same time 
this flux will have implications on the gauge couplings 
unification~\footnote{For such effects see for example~\cite{Blumenhagen:2008aw,Leontaris:2009wi,Leontaris:2017vzz}.}
 and the zero-mode  multiplicities of the spectrum
on the various matter curves.  To quantify these effects 
we introduce the  symbol $ {\cal F}_1$ for the $U(1)_{\chi}$ flux parameter and consider the flux 
restrictions on the matter curves
\ba
P={\cal F}_1\cdot (\chi-c_1);\; 
P_n={\cal F}_1\cdot (\chi_n-c_1);\; n=5,7;\; M={\cal F}_1\cdot (\eta-3c_1);\; 
C = -{\cal F}_1\cdot c_1~.
\ea

In this way we obtain the results shown in  the  last column of Table~\ref{T_ALL}.
We should mention that if we wish to protect the $U(1)_{\chi}$ boson from
receiving a Green-Schwarz (GS) mass we need to impose
\[ {\cal F}_1\cdot \eta =0\; \&\; {\cal F}_1\cdot c_1=0~,\]
which automatically imply $M=C=0$.
In this case, the sum $P=P_5+P_7$ stands for the total flux permeating matter curves
while one can observe form Table~\ref{T_ALL} that the flux
vanishes independently  on the $\Sigma_{16}$ and $\Sigma_{ 10}$ matter curves (Table \ref{Table 2}).

Assuming that $M_{10}^a$  is the number of $ \underline{10}_{t_1+t_3}\in SO(10)$,
after the $SO(10)$ breaking we obtain the multiplicities
for flipped representations:

\begin{equation}
\underline{16}_1=\begin{cases}10_{t_1},\; M_1\\
\bar{5}_{t_1},\;M_1+P\\
1_{t_1},\;M_1-P
\end{cases},\;\;\;
\underline{16}_2=\begin{cases}10_{t_3},\;M_3\\
\bar{5}_{t_3},\;M_3-P_5\\
1_{t_3},\;M_3+P_5
\end{cases},\;\;\;
\underline{16}_3=\begin{cases}10_{t_4},\;M_4\\
\bar{5}_{t_4},\;M_4-P_7\\
1_{t_4},\;M_4+P_7
\end{cases}
\end{equation}

\begin{equation}
10_1=\begin{cases}5^{(1)}_{-t_2-t_4},\;M_{10}^2\\
\bar{5}^{(1)}_{t_1+t_3},\;M_{10}^1+P
\end{cases},\;\;\;
10_2=\begin{cases}5^{(2)}_{-2t_1},\;M_{10}^1\\
\bar{5}^{(2)}_{t_3+t_4},\;M_{10}^1-P
\end{cases}
\end{equation}

\begin{table}[H]
	\begin{center}  \small%
		\begin{tabular}{|p{1cm}|p{1.0cm}|p{1.cm}|p{1cm}|p{1.0cm}| p{1.0cm}|p{1cm}|p{1cm}|}
			\hline
			$M_1$& $M_3$  & $M_4$ & $P$& $P_5$ &$P_7$ & $M_{10}^1$ & $M_{10}^2$ \\
			\hline
			$3$ & $1$ & $-1$ & $0$ & $1$ &$-1$ & $1$ & $0$\\
			\hline
		\end{tabular}
	\end{center}
	\caption{Model 1}\label{Table 2}
	\end{table}

\begin{align}
&10_{t_1}:3\times (Q_i,d_i^c,\nu_i^c),\;\; 10_{t_3}:1\times (H),\;\; 10_{t_4}:-1\times(\bar{H})\notag \\
&\bar{5}_{t_1}:3\times (u_i^c,L_i),\;\; 1_{t_3}:2\times (E_i^c),\;\; 1_{t_4}:-2\times (\bar{E}_i^c),\;1_{t_1}:3\times e^c_i\notag \\
&\bar{5}_{t_4+t_3}:1\times (\bar{h}),\;\; 5_{-2t_1}:1\times (h)~,
\end{align}

\noindent 
where $M_{10_i}, M_{5_j}$ stand for the numbers of $10\in SU(5)$ and $5\in SU(5)$
representations (a negative value corresponds to the conjugate representation).
$M_{S_{ij}}$ denote the multiplicities of the singlet fields. In fact,
as for any other representation, this means that
\be 
M_{ij}= \# 1_{t_i-t_j}-\# 1_{t_j-t_i},\label{NoSinglets}
\ee
thus, if $M_{ij}>0$ then there is an excess of  $M_{ij}$ singlets $1_{t_i-t_j}=\theta_{ij}$
and vice versa.

\section{The Superpotential and low energy predictions }

We will construct a model with  all three families residing 
on the same matter curve. Later on, we will explain  how in this case the masses to lighter families can be
generated by  non-commutative fluxes~\cite{Cecotti:2009zf}   or non perturbative effects~\cite{Aparicio:2011jx,Marchesano:2015dfa}.

Taking into account the  transformation properties of the various $SU(5)\times U(1)_{\chi}$ representations
presented in the previous section, we can readily write down the superpotential of the model. 
Regarding the  field content transforming non-trivially under $SU(5)\times U(1)_{\chi}$, we make the following identifications
\ba
10_{t_1}\to F_i,\; 
\bar 5_{t_1}\to \bar f_i,\; 1_{t_1}\to e^c_j,\; 1_{t_3}\to E^c_m,\; 1_{-t_4}\to \bar E^c_n,\\
10_{t_3}\to H, \,\overline{10}_{-t_4}\to \overline{H},\;
5_{-2t_1}\to h,\; \bar 5_{t_3+t_4}\to \bar h~.
\ea
Here the indices $i,j$ run over the number of families, i.e., $i,j=1,2,3$. 
All the representations emerging from the first matter curve labeled with 
$t_1$, share the same symbols as those of the field theory version of flipped $SU(5)$ of the previous section.   The two extra pairs with the quantum numbers of the right-handed electron
and its complex conjugate are denoted with $E^c, \bar E^c$.

Regarding the singlets $\theta_{pq},\; p,q=1,2,3,4$, taking into account the $Z_2$ monodromy $t_1\lra t_2$  we introduce the following naming:
\ba 
\theta_{12}\equiv \theta_{21}= s,\; \theta_{13}= \chi,\;\theta_{31}= \bar\chi,\; \theta_{14}\to  \psi, \;
\theta_{41}=  \bar \psi,\;\theta_{34}\to  \zeta,\;\theta_{43}\to \bar\zeta~.\label{singlets}
\ea 
The new symbols assigned to the $SU(5)$ massless spectrum of the
flipped model are collected in Table~\ref{T51}. A standard matter parity has also been assumed for all fields. 
 \begin{table}[H]
	\begin{center}
		\begin{tabular}{|lcc||ll|l|}
			\hline 
			Matter&& & Matter &&\\
			Field  &Symbol&Parity & Fields  &Parity&\\
			\hline
			$10_{-1} $&$F_i$&$-$&$\chi$&$+$&$M-P$
			\\
			$\bar 5_{3}$&$\bar f_i$&$-$&$\bar \chi$ &$+$&$P_5$
			\\
			$1_{-5}$&$e^c_i$&$-$&$\psi$&$+$ &$P_7$
			\\
			$1_0$&$s$&$-$&$\bar\psi$ &$+$&$M-P$
			\\
			$1_{5}$&$\bar E_n^c$&$-$ &$\zeta$  &$+$&$P$
			\\
			$1_{-5}$&$E_m^c$&$-$&$\bar\zeta$ & $+$&$P$
			\\
			\hline 
			$5_{2}$&$h$&$+$&$H$&$+$&$P$
			\\
			$\bar 5_{-2}$&$\bar h$&$+$&$\ov{H}$&$+$&$P$
			\\
			\hline
		\end{tabular}
	\end{center}
	\caption{ The $SU(5)\times U(1)_{\chi}$ representations with  their $R$-parity assignment. Their multiplicities are counted by the integers $M,P,P_{5,7}$ in the last column. }
	\label{T51}
\end{table}

Note that due to $t_1\lra t_2$ identification after the monodromy action, both types of singlets, $\theta_{12}$ and $ \theta_{21}$, are identified with the same one denoted with $s_j$,  with a  multiplicity $j=1,2,\dots, n_s$ determined by~(\ref{NoSinglets}).  For $M_{ij}=0$ there is an equal number of $\theta_{12}$ and $ \theta_{21}$
fields and  large mass terms of the form $M_{s_{ij}}s_is_j$ for all $s_i$ are normally expected. However, for $M_{ij}\ne 0$ some singlets are not expected  to receive tree-level masses.
 Such  `sterile' singlets $s_j$, (denoted collectively with $s$ in the following)  will play a significant role 
in relation to neutrino sector. Clearly, in addition to this, several other identifications will take place among the various flipped representations and the Yukawa couplings.  As an example, implementing the $Z_2$ monodromy and
the above definitions, the following gauge invariant terms are rewritten
as 
\ba 
10_{t_1}\bar 5_{t_2}\bar 5_{t_3+t_4}\xrightarrow{Z_2} 10_{t_1}\bar 5_{t_1}\bar 5_{t_3+t_4} &\to& F_i \bar f_j \bar h
\\
\ov{10}_{-t_4}10_{t_1} \theta_{21} \theta_{42}\xrightarrow{Z_2}\ov{10}_{-t_4}10_{t_1} \theta_{21} \theta_{41}&\to& \ov{H} F_i s\bar\psi ~.
\ea

With this notation the  superpotential terms are written in the familiar
field theory notation as follows:
\ba 
{\cal W}&=& \lambda_{ij}^u F_i\bar f_j \bar h+\lambda_{ij}^d F_iF_j h + \lambda_{ij}^e e^c_i \bar f_j h  + \kappa_i\overline{H} F_i s \,\bar\psi \nn\\
&&+ \alpha_{mj} \bar E^c_m e^c_j\,\bar \psi +\beta_{mn} \bar E^c_mE^c_n \,\bar \zeta+\gamma_{nj}E^c_n\bar f_j h\chi~. \label{superpot}
\ea 
The first three terms provide Dirac masses to the charged fermions and the neutrinos. It can be observed that the up-quark Yukawa coupling ($\propto F\bar f \bar h)$ appears at tree-level, as well as the bottom and charged lepton Yukawa couplings. Because in this construction  $U(1)_Y$ fluxes are not turned on, there is no splitting of the $SU(5)$ representations and thus, their corresponding content of the three generations resides on the  same matter curve. Using the geometric structure of the theory  it is possible to generate the fermion mass hierarchies and the Kobayashi-Maskawa mixing.    Here we give a brief account of  the mechanism, while the details are described in a considerable amount of work devoted to this issue~\cite{Heckman:2008qa,Heckman:2009de,Leontaris:2010zd,Camara:2011nj,Marchesano:2016cqg,Cvetic:2019sgs}.

We first recall that 
chiral matter  fields   reside on matter-curves  at the intersections of the GUT surface with other 7-branes, while the corresponding  wavefunctions, dubbed here $\Psi_i$, can be determined by solving the appropriate equations~\cite{Heckman:2009de} where it is found that they have a gaussian profile 
along the directions transverse to the matter-curve.
The tree-level superpotential terms of matter fields are formed at triple intersections  and each Yukawa coupling coefficient is determined by integrating over the overlapping wavefuctions
\[ \lambda_{ij} \propto \int_M \Psi_i \Psi_j \Phi_H dz_1\wedge d\bar z_1\wedge dz_2\wedge d\bar z_2~,\]
where $\Phi_H$ is the wavefucntion of the Higgs field.
Detailed computations of the Yukawa couplings with matter curves supporting the three generations, have shown that  hierarchical Yukawa matrices -reminiscent of the Froggatt-Nielsen mechanism-
are naturally obtained~\cite{Heckman:2008qa,Leontaris:2010zd,Camara:2011nj,Marchesano:2016cqg,Cvetic:2019sgs} with eigenmasses and mixing in agreement with the experimental values.

Returning to the superpotential terms~(\ref{superpot}), when the Higgs fields $\bar H$ and the singlet $\bar\psi$ acquire  non-vanishing VEVs, the last term of the first line in particular, generates a mass term coupling the right-handed neutrino with the singlet 
field $s$~\footnote{In order to simplify the notation, occasionally  the powers of $1/M_{str}^n$ (where $M_{str}$ is of the order of the string scale) in the non-renormalizable terms will be omitted. Hence we will write  $\bar\psi $ instead of $\bar\psi/M_{str}$ and so on.} :
\[ \kappa_i\langle \overline{H}\rangle \langle\bar\psi\rangle  F_i s  \,=\, M_{\nu_i^cs} \nu^c s~,\]
where $M_{\nu_i^cs}=  \kappa_i\langle \overline{H}\rangle \langle\bar\psi\rangle$. Bearing in mind that the top Yukawa coupling 
also implies a $3\times 3$ Dirac mass for the neutrino $m_{\nu_{D}}= \lambda_{ij}^u \langle \bar h\rangle$, and taking into account a  mass term  $M_s s s$ allowed by the symmetries of the model,  the following neutrino mass
matrix emerges
\be 
{\cal M}_{\nu}=\left(\begin{array}{ccc}
	0&m_{\nu_D}&0\\
	m^T_{\nu_D}&0&M^T_{\nu^cs}
	\\
	0&M_{\nu^cs}&M_s
\end{array}\right)~,\label{NMatrix}
\ee 
whereas additional non-renormalizable terms are also possible.
The low energy implications on various lepton flavor and lepton number violating processes  will be analysed  in section 6. Furthermore, the following terms  are also  consistent with the symmetries of the model:
\be 
{\cal W}\supset \lambda_{\mu}\chi\left(\psi+\overline{H} H \chi\right)\, \bar h \,h + \lambda_{\bar H} \overline{H} \overline{H} \bar h \bar\zeta 
+\lambda_{H} H H h  (\chi^2 +\bar\zeta^2\psi^2)~.
\ee 
When the various singlets acquire non-zero VEVs  the following fields receive masses. The  term proportional to $\lambda_{\mu}$ contains a non-renormalizable term proportional to $\chi\psi$ and a higher order one generated by the VEVs of Higges $\overline{H}H$. The terms proportional to $\lambda_{\bar H}, \lambda_{H}$ must provide   heavy  masses to the extra color triplet pairs  
$$\lambda_{\bar H}\langle  \overline{H}\rangle \dfrac{\langle \bar{\zeta} \rangle}{M_{str}}\,\ov{D}_{\bar H}^c \ov{D}_h 
+ \lambda_{ H}\langle H \rangle  \left(\dfrac{\langle\chi^2 \rangle}{M_{str}^2}+\dfrac{\langle \bar{\zeta}^2 \psi^2 \rangle}{M_{str}^4}\right) \,{D}_{\bar H}^c D_h~.$$
Since the magnitude of  $\langle \chi\rangle $  is constrained from the size of the  $\mu$ term,  large mass for the 
second triplet pair requires a large VEV for $\langle \psi\bar\zeta\rangle $. The solution of the flatness conditions in the appendix show that this is possible~\footnote{One might think that it would be possible to  eliminate the term $\chi \bar h h$ while  keeping the $\bar H \bar H \bar h$ and $H H h \bar\zeta \chi$ terms, by choosing appropriate $Z_2$ parity assignments  for $\chi$ and the other fields.
	It can be easily shown, however, that there is no such $Z_2$ assignment and possibly   generalized $Z_N$  or more involved symmetries are  required. Such discrete symmetries are available either from the spectral cover~\cite{Antoniadis:2013joa}, or from the torsion part of the Mordell-Weil group. }. According to the solution for flatness conditions problem obtained in the appendix, the useful singlets $\bar{\zeta},\bar{\psi},\chi$ acquire the desirable VEVs shown at  Table~\ref{Vevs}, generating this way an acceptable $\mu$-term for the Standard Model Higgs fields.
	\begin{table}
		\begin{center}  \small%
			\begin{tabular}{|p{1.5cm}|p{1.5cm}|p{1.5cm}|p{1.5cm}|p{1.5cm}| p{1.5cm}|p{1.5cm}|p{1cm}|}
				\hline
				$\chi$& $\bar{\chi}$  & $\psi$ & $\bar{\psi}$& $\zeta$ &$\bar{\zeta}$ \\
				\hline
				$5.6\times 10^{10}$ & $7.7\times 10^{15}$ & $2.2\times 10^{7}$ & $89.3\times 10^3$ & $7.8\times 10^{15}$ &$4.4\times 10^{15}$ \\
				\hline
			\end{tabular}
		\end{center}
		\caption{Masses in ${\rm GeV}$ scale.   $M_{str}=M_{GUT}=1.4\times 10^{16}\; {\rm GeV}$.}\label{Vevs}
	\end{table}

Continuing with the color triplet fields, we now collect all mass terms  derived from non-renormalizable contributions to the superpotential. They generate a $2\times 2$ mass matrix which is shown in Table~\ref{Triple}.

\begin{table}
\begin{center} \small%
\begin{tabular}{p{1cm}|p{3.5cm} p{3cm}}
$M^2_{D_h}$ & $D^c_H$ & $\overline{D_H^c}$\\
\hline
$D_h$ & $\langle H\rangle (\dfrac{\chi^2}{M_{str}^2}+\dfrac{\psi^2\bar{\zeta}^2}{M_{str}^4})$ & $\langle H\bar{H}\rangle (\dfrac{\chi^2}{M_{str}^3})$\\ \hline
$\overline{D_h}$ &$\langle H\bar{H}\rangle (\dfrac{\chi^2}{M_{str}^3})$ & $\langle \overline{H}\rangle \dfrac{ \overline{\zeta}}{M_{str}}$
\end{tabular}
\caption{The mass matrix for the down-type colour triplets.}
\label{Triple}
\end{center}
\end{table}

The Higgs color triplets  mediate baryon decay processes through dimension-four, and dimension-five  operators, thus their mass scale  is of crucial importance.
 Their eigenmasses  are 
\begin{align}
m_{D^c_H}=\langle H\rangle (\dfrac{\chi^2}{M_{str}^2}+\dfrac{\psi^2\bar{\zeta}^2}{M_{str}^4})\cos^2(\theta)&-\langle H\bar{H}\rangle (\dfrac{\chi^2}{M_{str}^3})\sin(2\theta)+\langle \overline{H}\rangle \dfrac{ \overline{\zeta}}{M_{str}}\sin^2(\theta)\notag \\
m_{\overline{D^c_H}}=\langle \overline{H}\rangle \dfrac{ \overline{\zeta}}{M_{str}}\cos^2(\theta)+\langle H\bar{H}\rangle (&\dfrac{\chi^2}{M_{str}^3})\sin(2\theta)+\langle H\rangle (\dfrac{\chi^2}{M_{str}^2}+\dfrac{\psi^2\bar{\zeta}^2}{M_{str}^4})\sin^2(\theta)\notag ~,
\end{align}
where the mixing angle $\theta$ is determined by
\begin{align}
\tan(2\theta)=&\dfrac{2\langle \bar{H} \rangle \langle \chi^2 \rangle M_{str}}{\langle \chi^2\rangle  M_{str}^2+\langle \psi^2\bar{\zeta^2}\rangle }~.
\end{align}

For  singlets VEVs of the order $10^{-1}M_{GUT} $,  the  triplets acquire heavy masses in the range  $10^{14}$-$ 10^{15}$\; GeV,  ($\theta \sim \frac{\pi}{6}$), protecting this way the proton from fast decays. For completeness, we summarize the possible proton decay processes in the next section.

\section{Proton Stability}




Having determined the masses of the color triplet fields $D,\bar D$, we are now able to examine possible bounds on the parameter space from proton decay processes. 
After the spontaneous breaking of the flipped $SU(5)$ gauge group, the resulting MSSM Yukawa Lagrangian contains   $B$ and $L$ violating  operators 
giving rise to proton decay  channels\cite{Ibanez:1991pr} such as $p\rightarrow (\pi^0,K^0) e^+$. Focusing our attention on the dangerous dimension five operators, in particular, the main contribution comes from the two relevant couplings $F_i F_j h,\; F_i\bar{f}_j\bar{h}$ in the superpotential \eqref{superpot}. Also, it is important to mention that color triplets can contribute through chirality flipping (LLLL and RRRR) operators and chirality non-flipping (LLRR) ones. 
 Following \cite{Ellis:2020qad,Hamaguchi:2020tet,Mehmood:2020irm}, these operators could be expressed in the mass eigenstate basis:
 \begin{align}
&10_{t_1}:(Q,VPd^c,U_{\nu^c}\nu^c),\;\; Q=(u,VPd)\notag\\
&\bar{5}_{t_1}:(u^c,U_{L}L),\;\; L=(U_{\small{PMNS}}\nu,e)\notag\\
&1_{t_1}:(U_e e^c)~.
\end{align}
Therefore, the color triplets couplings  to ordinary  MSSM matter fields are expressed as
\begin{align}
 \lambda_{ij}^u&: Q(V^* \lambda^{(d^c)} V^{\dagger})Q D^c_H\notag\\
 \lambda_{ij}^e&: u^c (U_L^{\dagger} \lambda^{(e^c)})e^c D^c_H \notag \\
 \lambda_{ij}^u&: L(U_L \lambda^{(Q,\nu)})Q \overline{D^c_H}\notag \\
 \lambda_{ij}^u&: u^c( \lambda^{(Q,\nu)}V)d^c \overline{D^c_H},
\end{align}
where $V$ is the   Cabbibo-Kobayashi-Maskawa (CKM) matrix with the corresponding phases and $U_L$ is the leptonic part of the $PMNS$-matrix $U_{PMNS}=U_L^*U_{\nu}^{\dagger}$, plus the CP-phases $P={\rm diag}(e^{i\phi_i})$.
The dominant effects on proton decay originate from LLRR channels, where after integrating out the Higgs triplets (recall that in this diagram chirality flipped dressing with a higgsino is required), are discussed below. These operators, also, should respect the $SU(4)_{\bot}$ charge conservation, so for each operator the appropriate singlet  fields must be introduced. Since the masses of these singlets are substantially lower that the string scale,  further suppression of the anticipated baryon violating  operators is expected. The relevant operators take the form
\begin{equation}
	\begin{split}
			\delta_1\dfrac{10_{t_1} 10_{t_1}10_{t_1}\bar{5}_{t_1}}{M_{str}}(\dfrac{\theta_{31}\theta_{41}}{M_{str}^2}+\dfrac{\theta_{31}^2\theta_{43}}{M_{str}^3})&
	\rightarrow \delta_1\dfrac{\langle \bar{\chi}^2\bar{\zeta}\rangle +\langle \bar{\chi}\bar{\zeta}\rangle M_{str}}{M_{str}^4}(Q_i Q_j Q_k L_m)
	\\
	\delta_2\dfrac{10_{t1} \bar{5}_{t_1}\bar{5}_{t_1}1_{t_1}}{M_{str}}(\dfrac{\theta_{31}\theta_{41}}{M_{str}^2}+\dfrac{\theta_{31}^2\theta_{43}}{M_{str}^3})&\rightarrow \delta_2\dfrac{\langle \bar{\chi}^2\bar{\zeta}\rangle +\langle \bar{\chi}\bar{\zeta}\rangle M_{str}}{M_{str}^4}(d_i^c u_j^c u_k^c e^c_m),
	\end{split}
\end{equation}
where $\delta_{1,2}$ are 
\begin{equation}
\delta_1\sim \dfrac{\langle h \rangle}{m_{D^c_H}m_{\overline{D^c_H}}}\big[(V^{*}\lambda^{(d^c)}V^{\dagger})(\lambda^{(Q,\nu)}U^*_L)\big],\;\; \delta_2\sim  \dfrac{\langle h \rangle}{m_{D^c_H}m_{\overline{D^c_H}}}\big[(U_L^{*}\lambda^{(e^c)})(\lambda^{(Q,\nu)}V)\big]~.
\end{equation}
Given the scale difference between the bidoublet $\langle h\rangle$ and the triplet $ M_{D_H}^c $, these operators are highly suppressed. The novelty of F-theory model building constructions compared to GUT-model building~\cite{Hamaguchi:2020tet,Mehmood:2020irm}, is that the  $t_i$-charge conservation implies additional suppression. Regarding the chirality flipping diagrams, as it is pointed out in~\cite{Hamaguchi:2020tet}, they are severely constrained in the flipped $SU(5)$ model, as opposed to their behavior in the standard $SU(5)$ \cite{Sakai:1981pk}.

\par We investigate now the implications of the various  dimension-6 operators.  In this case, baryon violating  decays are mediated by both $SU(5)$ vector gauge fields and color Higgs triplets. The corresponding  diagrams differ from dimension five operators, since chirality flipping is not needed in this case, so the extra suppression factor  $\frac{\langle h \rangle}{M_D}$ is absent. From the low energy superpotential~\eqref{superpot}, the relevant to  proton decay couplings are:
\begin{equation}
 \lambda_{ij}^u F_i\bar f_j \bar h+\lambda_{ij}^d F_iF_j h\,\bar\psi + \lambda_{ij}^e e^c_i \bar f_j h \,\bar \psi ~,
\end{equation}
whereas, the effective operators corresponding to dimension-6 operators are:
\begin{equation}
10\; \bar{5}\; 10^{\dagger}\;\bar{5}^{\dagger},\;\; 10\; 10\; \bar{5}^{\dagger}\;1^{\dagger}\notag~.
\end{equation}


\noindent The gauge interactions inducing the dimension six operators can be summarized as:

\begin{equation}
\mathcal{L}\sim g_5\bigg( \epsilon_{ij} u^c X^i U_L^* L^j+\epsilon_{abc}Q^{\dagger a}X^b VP^*d^c+\epsilon_{\alpha \beta}\nu^{\dagger c}X^{\alpha}Q^{\beta}+h.c.\bigg)~,
\end{equation}
and
\begin{equation}
\mathcal{L}_{(6)}\sim C_{(6)\alpha}^{ijkm}\;\bigg(u_i^{\dagger c} d_j^{\dagger c} (u_k e_m+ d_k\nu_m)\bigg)+C_{(6)\beta}^{ijkm}\; \bigg(u_i(VP^*d_j)+(V^*Pd_i)u_j\bigg)u_k^{\dagger c}e_m^{\dagger c}~.
\end{equation}
The coefficients $C_{(6)\alpha,\beta}^{ijkm}$ are given by~\cite{Hamaguchi:2020tet,Mehmood:2020irm}

\begin{align}
C_{(6)\alpha}^{ijkm}=&\bigg( \dfrac{(U_L)_{km}V^*_{ij}}{M^2_G}+\dfrac{(V^{\dagger}\lambda^{(Q,\nu)})_{ij}(U_L\lambda^{(Q,\nu)}_{km})}{m_{\overline{D^c_H}}^2}\bigg)\notag \\
C_{(6)\beta}^{ijkm}=&\bigg( \dfrac{(V^*P\lambda^{(d^c)}V)_{km}(U^{\dagger}_L\lambda^{(e^c)})_{ij}}{m_{D^c_H}^2} \bigg)~,
\end{align}
where $M_G$ is the mass of the gauge boson and the Yukawa couplings $\lambda$ are the diagonal matrices. It is important to emphasize that the flipped $SU(5)$ gauge bosons do not couple to the right-handed leptons, in contrast to the standard $SU(5)$. The final state is different in these two cases and their experimental implication makes the flipped version much more phenomenologically attainable (see also~\cite{Ellis:2020qad}).  As an illustrative example,  we  present the charged lepton decay channels $p\rightarrow (K^0,\pi^0) l_{(e,\mu)}^+$. First of all the mixing factors, for the two Wilson coefficients stated above, are:
\begin{align}
p\rightarrow \pi^0 l_i^+&:\;\; (U_L)_{i1}V^*_{ud}(e^{\phi_{u}},e^{\phi_d})\notag\\
p\rightarrow K^0 l_i^+&:\;\; (U_L)_{i1}V^*_{us}(e^{\phi_{u}},e^{\phi_s}),
\end{align}
where the  index $i$ denotes the generation of the lepton involved in the proton decay. The decay rates can be computed as:

\begin{align}
\Gamma_{p\rightarrow \pi^0 e^+}&=  |(U_L)_{11}V^*_{ud}(e^{\phi_{u}},e^{\phi_d})|^2 \mathcal{K}(m_{\pi},m_{p})\mathcal{M}^2(\pi^0,e^{+})\bigg[A_{\alpha}^2(\dfrac{1}{M_G^2}+\dfrac{f^2(u)}{m_{\overline{D^c_H}}^2})^2+A_{\beta}^2(\dfrac{g^2(d,e^+)}{m_{D^c_H}^2})^2 \bigg],\notag\\
\Gamma_{p\rightarrow K^0 e^+}&=  |(U_L)_{11}V^*_{us}(e^{\phi_{u}},e^{\phi_s})|^2 \mathcal{K}(m_{K^0},m_{p})\mathcal{M}^2(K^0,e^{+})\bigg[A_{\alpha}^2(\dfrac{1}{M_G^2}+\dfrac{f^2(u)}{m_{\overline{D^c_H}}^2})^2+A_{\beta}^2(\dfrac{g^2(s,e^+)}{m_{D^c_H}^2})^2 \bigg],\label{Proton}
\end{align}
where $A_{\alpha},A_{\beta}$ are the renormalization factors obtained from the RGE equations (in one-loop level) for the Wilson coefficients contributing to the proton decay processes \cite{Ellis:2020qad,Hamaguchi:2020tet,Mehmood:2020irm}. Since there are some additional states in the low energy spectrum (namely the vector-like singlets $E^c$), we do not expect a significant deviation for the gauge coupling unification regarding the supersymmetry (susy) breaking scale around TeV, as obtained by similar analysis~\cite{Jiang:2006hf}. The rest of the parameters used in the decay rates are summarized below:

\begin{align}
\mathcal{K}(m_{\pi},m_{p})&=\dfrac{m_p}{32\pi}\big(1-\dfrac{m_{\pi^0}^2}{m_p^2}\big)^2,\;\;\mathcal{M}(\pi^0,(e^{+},\mu^+))=\langle \pi^0|(ud)_{R}u_L|p\rangle_{l^+}=(-0.131,-0.118) \; {\rm GeV}^2,\notag\\
\mathcal{K}(m_{K^0},m_{p})&=\dfrac{m_p}{32\pi}\big(1-\dfrac{m_{K^0}^2}{m_p^2}\big)^2,\;\mathcal{M}(K^0,(e^{+},\mu^+))=\langle \pi^0|(us)_{R}u_L|p\rangle_{l^+}=(0.103, 0.099) \; {\rm GeV}^2,\notag\\
f^2(u)=&\dfrac{m^2_u}{\langle h_u \rangle^2}, \;\;g^2(d, e^+)=\dfrac{m_u m_{e^+}}{\langle h_d \rangle^2}, \;\;g^2(s,e^+)=\dfrac{m_s m_{e^+}}{\langle h_d \rangle^2}, \; \tan(\beta)=\dfrac{\langle h_u \rangle }{\langle h_d \rangle}~.
\end{align}

\begin{figure}[H]
    \centering
    \begin{minipage}{.5\textwidth}
        \centering
        \includegraphics[scale=0.9]{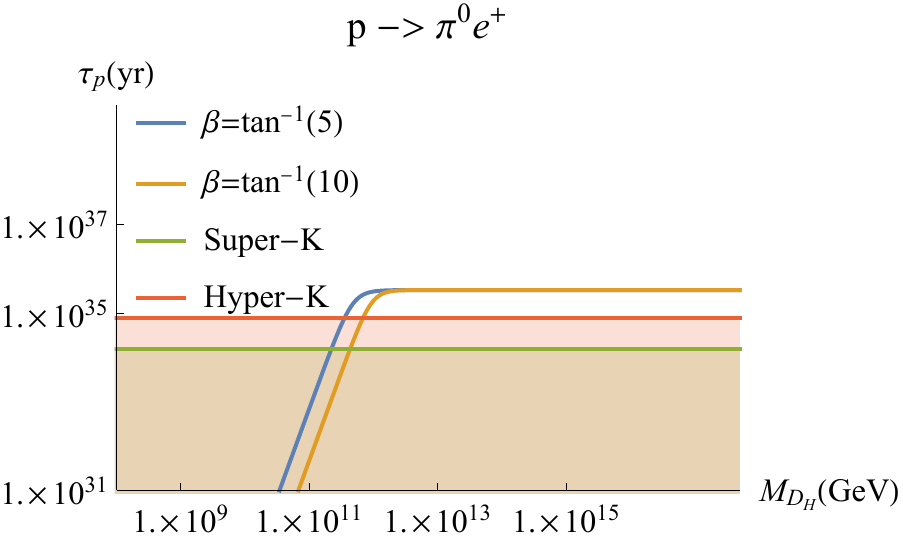}
    \end{minipage}%
    \begin{minipage}{0.7\textwidth}
        \centering
        \includegraphics[scale=0.9]{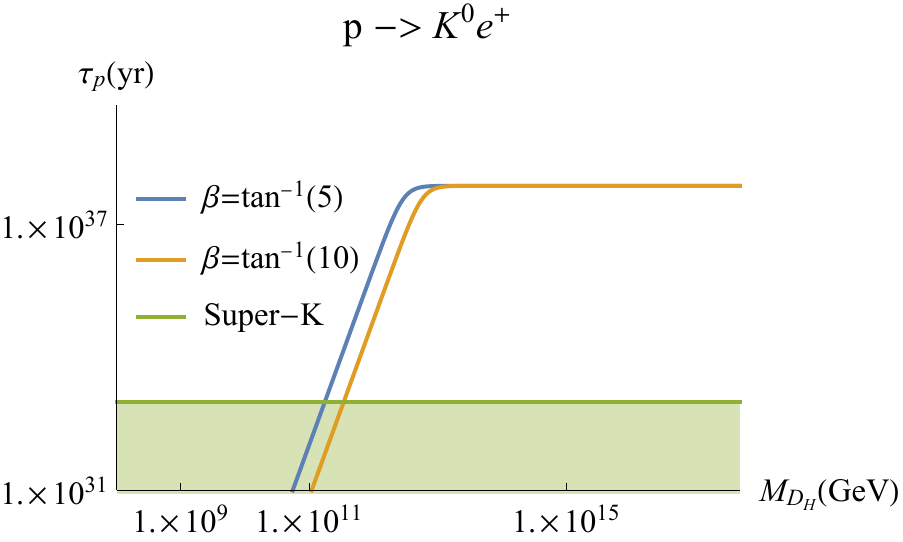}
    \end{minipage}
    
\centering
    \begin{minipage}{.5\textwidth}
        \centering
        \includegraphics[scale=0.9]{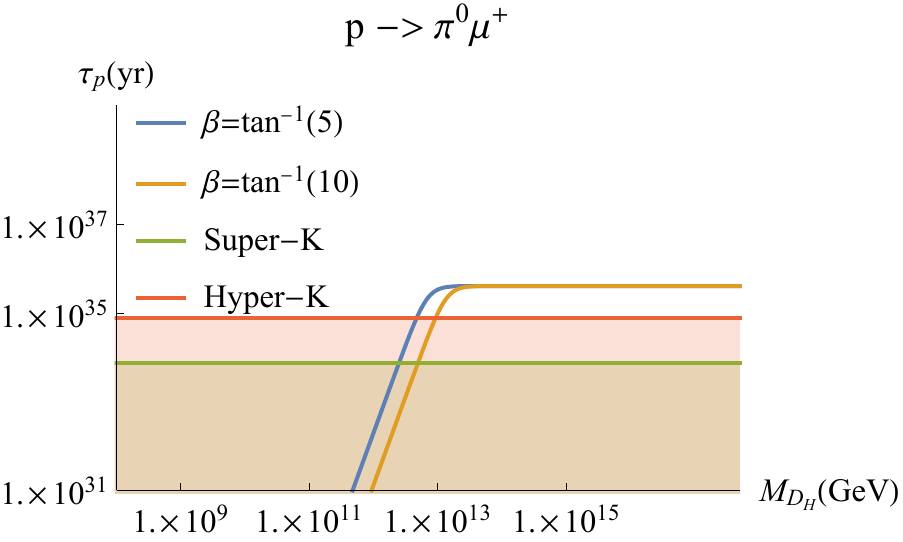}
    \end{minipage}%
    \begin{minipage}{0.7\textwidth}
        \centering
        \includegraphics[scale=0.9]{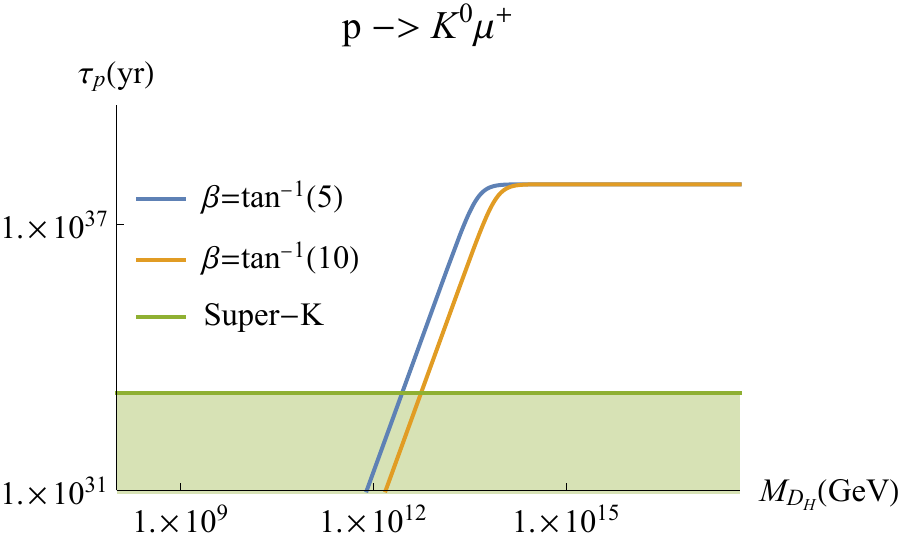}
    \end{minipage}

    \caption{The lifetime of the proton along the two decay channels ($p\rightarrow \pi^0(e^+,\mu^{+})$, $p\rightarrow K^0(e^+\mu^{+})$) for different values of $\tan(\beta)$. It is  deduced that the triplets mass is bounded at $m_{\overline{D^c_H}}=m_{D^c_H}\geq 10^{11}\; {\rm GeV}, M_G=10^{16}\; {\rm GeV}$. The asymptotic value of the lifetime is controlled by the masses of the Higgs triplets.}\label{pdec}
\end{figure}

In figure~\ref{pdec} we plot the proton lifetime of the above decay channels,  as a function of 
the triplet mass $m_{D_H}$ for assuming various values of $\tan\beta$, where the horizontal lines represent the current Super-K~\cite{Super-Kamiokande:2016exg} and Hyper-K~\cite{Hyper-Kamiokande:2018ofw} bounds. Regarding the formulas for the proton decay through the muon's channel, they can be easily derived if we trade the $e^+\rightarrow \mu^+$.

 \section{The Neutrino Sector}

In this section  we are going to examine in some detail the mass matrix \eqref{NMatrix} involving the neutrinos and the 
neutral singlet fields $s$. Recall that the latter are identified with the singlets $\theta_{12},\theta_{21}$ and that their number is determined by global dynamics of the model. In the present semi-local  construction we will treat them as a free parameter.   The  following Yukawa  couplings
\begin{equation}
m_{\nu_D}= \lambda_{ij}^u \langle\bar h \rangle,\;\;M_{\nu_i^cs}=  \dfrac{\kappa_i\langle \overline{H}\rangle \langle\bar\psi\rangle}{M_{str}}~,
\end{equation}
define the Dirac neutrino mass submatrix and the mixing between the right-handed neutrinos and the singlet fields. 
Additional non-renormalizable terms may also generate masses for the right-handed neutrinos $\nu_i^c$ due to a coupling of the form :
\begin{eqnarray}
\mathcal{W}&\sim& \dfrac{\lambda_{ij}}{M_{str}^3}\overline{H}\overline{H}F_{i}F_{j}\left(\langle\bar{\psi}^2\rangle+\dfrac{\langle \bar{\zeta}\rangle^2\langle\bar{\chi}\rangle^2}{M_{str}^2}\right) \;\Rightarrow \nonumber 
\\
 M_{\nu_i^c}&=&\dfrac{\lambda_{ij}\langle \bar{\nu}_H^c\rangle^2}{M_{str}^3}\left(\langle\bar{\psi}^2\rangle+\dfrac{\langle \bar{\zeta}\rangle^2\langle\bar{\chi}\rangle^2}{M_{str}^2}\right) ~.
\end{eqnarray}

Hence, the final structure of the neutrino mass sector is
\be 
{\cal M}_{\nu}=\left(\begin{array}{ccc}
0&m_{\nu_D}&0\\
m^T_{\nu_D}& M_{\nu_i^c} &M^T_{\nu^cs}
\\
0&M_{\nu^cs}&M_s
\end{array}\right)~.\label{NMatrix1}
\ee 

 This matrix involves vastly different scales. We assume (also justified by the singlet VEVs) the hierarchy $m_{\nu_D}\ll M_s \ll  M_{\nu_i^cs},M_{\nu_i^c}$ and  implement a double inverse seesaw mechanism to determine the eigenvalues of the light spectrum. Below we sketch the procedure for obtaining the normal-order mass hierarchy in the light neutrinos sector.
We define:

\begin{align}
M_{\nu_D}=\begin{pmatrix}
m_{\nu_D}\\
0
\end{pmatrix},\;\; M_{R^{'}}=\begin{pmatrix}
M_{\nu_i^c} & M^T_{\nu^cs}\\
M_{\nu^cs}  & M_s
\end{pmatrix}~,
\end{align}
and 
\begin{align}
M_{\nu}=\begin{pmatrix}
0 & M_D^T\\
M_D & M_{R^{'}}\label{nmatrix}
\end{pmatrix}~.
\end{align}
 Then, implementing the double inverse seesaw formula (see for example~\cite{Zhou:2012ds}) we obtain
\begin{align}
m_{\nu_i}=&-m_{\nu_D}(M_{\nu_i^c}-M_{\nu^cs} M_s^{-1} M^T_{\nu^cs})^{-1}m^T_{\nu_D}\notag \\
&m_{\nu_D}\ll (M_{\nu_i^c}-M_{\nu^cs} M_s^{-1} M^T_{\nu^cs})~.
\end{align}
Depending on the scale of the neutral singlets $s$,  there are two basic limits of the previous equation, which yield different parametric regions for the right-handed neutrinos and the singlets. 
In the subsequent sections we would like to implement a leptogenesis scenario, hence it is of crucial importance to pursue an intermediate mass scale ($\sim$ TeV) in the heavy neutrinos sector and to characterize the properties of the extra singlets.  Having this in mind, we proceed with 
the analysis of the limiting cases.
 \par 
$\alpha$)  We assume the hierarchies $M_{\nu_i^c}\ll M_{\nu^cs}$ and $M_s\ll M_{\nu^cs}$.\\
In this case, the $\{22\}$-entry in the neutrino mass matrix is less significant and the model reduces to the standard double seesaw:    

\begin{equation}
m_{\nu_i}=m_{\nu_D}{(M^{T}_{\nu^cs})}^{-1} M_s M^{-1}_{\nu^cs}m^T_{\nu_D}~.
\end{equation}
This scenario accommodates effectively the light neutrino masses, where for example requiring light neutrinos at sub-eV scale $m_{\nu_i}\lesssim 0.1\; eV$ and sterile masses around $M_s\sim 5\; keV$ ($m_{\nu_D}\sim 100\; GeV$), the seesaw scale for the right-handed neutrinos is set at $M_{\nu^cs}\sim TeV$.  A much more interesting and testable prediction from such a case would be the calculation of unitarity violation $\eta$ in the leptonic mixing matrix \cite{Hettmansperger:2011bt}:

\begin{equation}
V=(1+\eta)U_{0},\label{etadef}
\end{equation}
where the $V$ matrix diagonalizes the light neutrinos and $U_{0}$ represents the unitary matrix (identified with $U_{PMNS}$ in the lepton sector), while the $\eta$ matrix can in principle be hermitian. Deviations from the unitary form of the PMNS mixing matrix are displayed into the rare leptonic decays ($l_a\rightarrow l_b \gamma$). These decays put stringent bounds on the discrepancies in the mixing matrix, whose origin can be traced back to the seesaw mechanism. In order the explain how deviations can be expressed, it is important to recall the GIM mechanism~\cite{Glashow:1970gm} . Flavor changing neutral currents are induced at loop level in the Standard Model, where their decay rate is parametrized in terms of the mixing matrix in 1-loop as \cite{Petcov:1976ff}: 

\begin{align}
&\dfrac{\Gamma(l_a\rightarrow l_b \gamma)}{\Gamma(l_a\rightarrow \nu_a l_b \bar{\nu}_b)}\sim \dfrac{|\sum_k V_{ak}V^{\dagger}_{kb}F(\frac{m_{\nu}^2}{m_{W}^2})|^2}{(VV^{\dagger})_{aa}(VV^{\dagger})_{bb}},\notag\\
F(x)&=\dfrac{10-43x+78x^2-49x^3+4x^4+18x^3\log(x)}{3(x-1)^4} ,
\end{align}

\noindent 
where for unitary mixing matrix $U$ the GIM mechanism implies a vanishing contribution for $a\neq b$ \cite{Antusch:2006vwa}. In the case of non-unitary mixing matrix, a typical process $\mu\rightarrow e\gamma$ results in the experimental bound $( U_{e\mu}U^{\dagger}_{\mu e})<10^{-4}$,which represents the typical condition needed to be met by seesaw scenarios. Regarding the computation of the unitary violating effects $\eta$, they can be computed by the neutrino matrix \eqref{nmatrix}, using the matrix \eqref{etadef}, as:

\begin{equation}
\eta\cong -\dfrac{1}{2}M_D^{\dagger}(M_{R}^*)^{-1}(M_{R})^{-1}M_D~.
\end{equation}

Regarding the unitarity violation in the seesaw mechanism analysed here, an estimate of the $\eta$ can be computed after the scales of the seesaw matrix are set. Nevertheless, in both of the two limits of the seesaw mechanism analyzed here, the $\eta$ parameter is of order:

\begin{equation}
\eta\sim \mathcal{O}(\dfrac{m^2_{\nu_D}}{{M^{2}_{\nu^cs}}}) \sim   10^{-6},
\end{equation}
i.e., two orders below the present bound.

\textit{$\beta$)} $M_s\ll M_{\nu^cs}\ll M_{\nu^c}$.   In this limit,
the two heavy states are
\begin{align}
\hat{m}_{s}=M_s&-M_{\nu^cs}^TM_{\nu^c}^{-1}M_{\nu^cs},\notag\\
\hat{m}_{\nu^c}&=M_{\nu^c}~.
\end{align}
Regarding the light neutrino states, depending on the heavy mass hierarchies, we distinguish two cases. 
For $M_{\nu^c}\ll M_{\nu^cs} M_s^{-1} M^T_{\nu^cs}$,

\begin{equation}
m_{\nu}=m_{\nu_D}{(M^{T}_{\nu^cs})}^{-1} M_s M^{-1}_{\nu^cs}m^T_{\nu_D},
\end{equation}
and for $M_{\nu^c}\gg M_{\nu^cs} M_s^{-1} M^T_{\nu^cs}$,
\begin{align}
m_{\nu}&=-m_{\nu_D}M_{\nu^c}^{-1}m^T_{\nu_D}~.\label{TwoCases}
\end{align}
 
In the first case, the  paradigm ($\alpha$) is reproduced and in the second one the typical seesaw is obtained. Here, the new intermediate scale $\tilde{m}_s$ could be useful for a dark matter particle, since the mixing angle between the active and the sterile neutrino is highly suppressed. This angle could be obtained after integrating out the heavy right-handed neutrino scale  $M_{\nu^c}$, leading to:
\begin{align}
\tan(2\theta_{\nu s})\cong&\dfrac{2m_{\nu_D}}{M_{\nu^cs}}, \;\;(M_{\nu_i^c},M_{s}\ll M_{\nu^cs})\; \&\;(M_{\nu^c}\ll M_{\nu^cs} M_s^{-1} M^T_{\nu^cs}),\\
\tan(2\theta_{\nu s})\cong &\dfrac{m_{\nu_D}M_{\nu^cs}}{2M_sM_{\nu^c}},\;\; M_{\nu^c}\gg M_{\nu^cs} M_s^{-1} M^T_{\nu^cs}
~.
\end{align}
The mixing angle of the active-sterile neutrinos are of crucial importance, since this angle characterizes the sterile neutrinos'  properties regarding its nature as a dark matter particle. Astrophysical data have already opened two ``windows'' for sterile dark matter  particles, the first one at keV scale with the mixing angle $\theta_{\nu s}\sim (10^{-6},10^{-4})$ and the second one at $MeV$ scale with $\theta_{\nu s}\sim (10^{-9},10^{-6})$.

\subsection*{Leptogenesis}
Next we examine  the leptogenesis scenario in the context of the  flipped $SU(5)$ model presented in this work. Our analysis shows that a possible implementation of the leptogenesis scenario can be realized in the second case (i.e., case $\beta$).  As is well known, right-handed neutrinos can decay to a lepton and a Higgs field, producing this way  lepton asymmetry.
The relevant Yukawa couplings are
\ba 
{\cal W}&=& \lambda_{ij}^u F_i\bar f_j \bar h  + \kappa_i^{\prime}\overline{H} F_i s \,\bar\psi \nn,\;\; \kappa^{\prime}_i=\kappa_i\dfrac{\langle \bar{\psi} \rangle}{M_{str}}~.
\ea 
Figure~2 shows the relevant vertex of the right-handed neutrino and the  standard  one-loop graph contributing to the lepton asymmetry. There are also two wavefucntion  self-energy one-loop correction graphs depicted in figure 3 which also contribute.

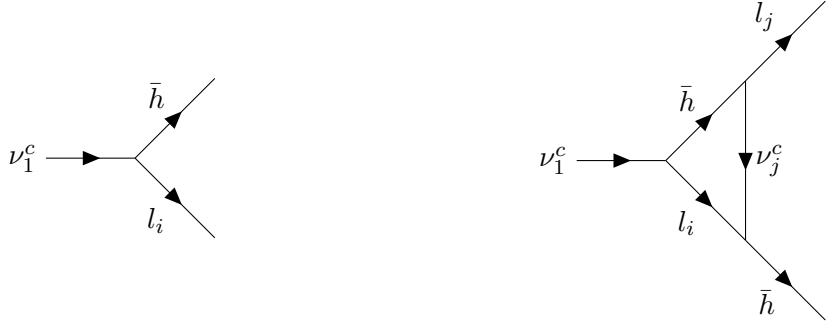
\begin{figure}[H]

\begin{subfigure}{0.5\textwidth}
\centering

\begin{tikzpicture}
  \begin{feynman}
    \vertex (a) {\(\nu_1^c\)} ;
    \vertex [right=of a] (b);
    \vertex [above right=of b](c);
    \vertex [below right=of b] (d);

    \diagram* {
      (a) -- [fermion] (b) -- [fermion,edge label=\(\bar{h}\)] (c),
      (b) -- [fermion, edge label'=\(l_i \)] (d),

    };
  \end{feynman};
\end{tikzpicture}
\end{subfigure}
\begin{subfigure}{0.5\textwidth}
\centering
\begin{tikzpicture}
  \begin{feynman}
    \vertex (a) {\(\nu_1^c\)} ;
    \vertex [right=of a] (b);
    \vertex [above right=of b](c);
    \vertex [below right=of b] (d);
    \vertex[below right=of d](f);
    \vertex[above right=of c](g);

    \diagram* {
      (a) -- [fermion] (b) -- [fermion,edge label=\(\bar{h}\)] (c),
      (b) -- [fermion, edge label'=\(l_i \)] (d),
     (d)-- [fermion, edge label'=\(\bar{h}\)](f),
     (c) -- [fermion, edge label=\(l_j\)](g),
     (c) -- [fermion, edge label=\(\nu_j^c\)](d)
 
    };
  \end{feynman};
\end{tikzpicture}
\end{subfigure}
\caption{Standard contributions to the generated lepton asymmetry.}
\end{figure}

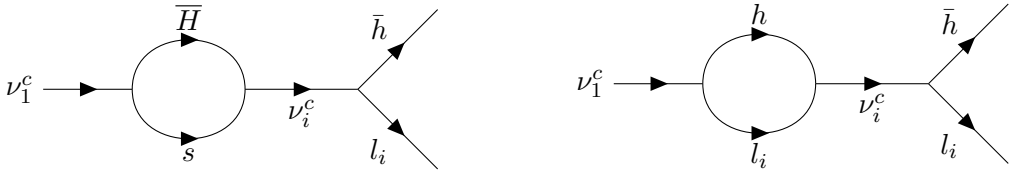
\begin{figure}[H]
\begin{subfigure}{0.5\textwidth}
\centering

\begin{tikzpicture}
  \begin{feynman}
    \vertex (a) {\(\nu_1^c\)} ;
    \vertex[right=of a](e);
    \vertex[right=of e](f);
    \vertex[right=of f](b);
    \vertex [above right=of b](c);
    \vertex [below right=of b] (d);

    \diagram* {
      (a)-- [fermion](e),
      (e)--[fermion, half left, edge label=\(\overline{H} \)](f),
       (e)--[fermion, half right, edge label'=\(s\)](f),
      (f) -- [fermion, edge label'=\(\nu_i^c\)] (b) -- [fermion,edge label=\(\bar{h}\)] (c),
      (b) -- [fermion, edge label'=\(l_i \)] (d),

    };
  \end{feynman};
\end{tikzpicture}
\end{subfigure}
\begin{subfigure}{0.5\textwidth}
\centering
\begin{tikzpicture}
  \begin{feynman}
    \vertex (a) {\(\nu_1^c\)} ;
    \vertex[right=of a](e);
    \vertex[right=of e](f);
    \vertex[right=of f](b);
    \vertex [above right=of b](c);
    \vertex [below right=of b] (d);

    \diagram* {
      (a)-- [fermion](e),
      (e)--[fermion, half left, edge label=\(\bar{h}\)](f),
       (e)--[fermion, half right, edge label'=\(l_i\)](f),
      (f) -- [fermion, edge label'=\(\nu_i^c\)] (b) -- [fermion,edge label=\(\bar{h}\)] (c),
      (b) -- [fermion, edge label'=\(l_i \)] (d),

    };
  \end{feynman};
\end{tikzpicture}
\end{subfigure}
\caption{Loop diagrams contributions to the generated lepton asymmetry.}
\end{figure}

The decay rate is given by
\begin{align}
\Gamma(\nu^c_i)=\dfrac{1}{4\pi}\bigg( \lambda_{ij}^{\nu} (\lambda_{ij}^{\nu})^{\dagger}+\kappa^{\prime} (\kappa^{\prime})^{\dagger}\bigg)_{ii}M_{\nu^c_i}~,
\end{align}
where $\lambda$ and $\kappa^{\prime}$ are the relevant Yukawa couplings in the equation \eqref{superpot} for the neutrino sector. The lepton asymmetry factor is summarized to the following contributions:

\begin{equation}
\epsilon_1=-\sum_i \dfrac{\Gamma_1(\nu_1^c\rightarrow \bar{l}_i \bar{h})-\Gamma_2(\nu_1^c\rightarrow l_i h)}{\Gamma_{12}(\nu_1^c)},
\end{equation}
\noindent 
where $\Gamma_{12}=\Gamma_1(\nu_1^c\rightarrow \bar{l}_i \bar{h})+\Gamma_2(\nu_1^c\rightarrow l_i h)$ indicates the overall decay rates. The lepton asymmetry in such a scenario can be written as~\cite{Covi:1996wh}:

\begin{align}
\epsilon_1=\dfrac{1}{8\pi}\sum_{j\neq 1}\bigg( (f_1(x_j)+f_2(x_j))&G_{j1}+f_2(x_j)G^{\prime}_{j1}\bigg)~,\\
f_1(x_j)=\sqrt{x}(1-(1+x)\ln(\frac{1+x}{x})),\;\; &f_2(x)=\frac{\sqrt{x_j}}{1-x_j}, \;\; x_j=\frac{M_{\nu^c_j}^2}{M_{\nu^c_1}^2},
\end{align}
where the $f$-factors are the vertex contributions of the Feynman diagrams. Now, the $G$-factors contain the Yukawa couplings as:
\begin{align}
G=\dfrac{Im\big[( \lambda_{ij}^{\nu} (\lambda_{ij}^{\nu})^{\dagger})^2\big]}{( \lambda^{\nu} (\lambda^{\nu})^{\dagger}+\kappa^{\prime} (\kappa^{\prime})^{\dagger})_{11}},\;\; G^{\prime}=\dfrac{Im\big[( \lambda_{ij}^{\nu} (\lambda_{ij}^{\nu})^{\dagger})(\kappa^{\prime}(\kappa^{\prime})^{\dagger})\big]}{( \lambda^{\nu} (\lambda^{\nu})^{\dagger}+\kappa^{\prime} (\kappa^{\prime})^{\dagger})_{11}}~.
\end{align}

 With regard to the impact of the loop corrections of the second graph in figure 3, the lepton asymmetry factor can be divided into two cases with respect to the right-handed neutrino mass hierarchy $x_j= \frac{M_{\nu^c_j}^2}{M_{\nu^c_1}^2}$.  For the case of large hierarchy, $x_j\gg 1$, the contribution from the loops is negligible resulting in \cite{Davidson:2002qv}:

\begin{align}
&\epsilon_1\cong -\dfrac{3M_{\nu^c_1}}{16\pi \langle v \rangle ^2}\dfrac{Im\big[(\lambda_{ij}^{\nu})^*m_{\nu}(\lambda_{ij}^{\nu})^{\dagger}\big]}{( \lambda^{\nu} (\lambda^{\nu})^{\dagger}+\kappa^{\prime} (\kappa^{\prime})^{\dagger})_{11}}\Rightarrow \notag\\
&|\epsilon_1|\lesssim \dfrac{3M_{\nu^c_1}}{16\pi\langle m_{\nu_D}\rangle^2}(m_{\nu_3}-m_{\nu_1})~.
\end{align}
From the above, it is obvious that in order to obtain the observed lepton asymmetry $\epsilon_1\sim [10^{-6},10^{-5}]$, the scale for the right-handed neutrinos should lay close to:
\begin{equation}
M_{\nu^c_1}\gtrsim \dfrac{16\epsilon_1\pi\langle m_{\nu_D}\rangle^2}{3(m_{\nu_3}-m_{\nu_1})}\gtrsim 10^9\; GeV~.\label{leptcase1}
\end{equation}

The  case  $x_j\cong 1$ describes the enhancement due to the loop diagrams (resonant procedure), where the asymmetry factor is:

\begin{equation}
\epsilon_1\cong -\dfrac{1}{16\pi}\left\{\dfrac{M_{\nu^c_2}}{\langle m_{\nu_D} \rangle^2}\dfrac{{\rm Im}[(\lambda_{ij}^{\nu})^*m_{\nu}(\lambda_{ij}^{\nu})^{\dagger}]}{( \lambda^{\nu} (\lambda^{\nu})^{\dagger}+\kappa^{\prime} (\kappa^{\prime})^{\dagger})_{11}}+\dfrac{\sum_{j\neq 1}{\rm Im}[( \lambda_{ij}^{\nu} (\lambda_{ij}^{\nu})^{\dagger})(\kappa^{\prime} (\kappa^{\prime})^{\dagger})]}{( \lambda^{\nu} (\lambda^{\nu})^{\dagger}+\kappa^{\prime} (\kappa^{\prime})^{\dagger})_{11}}\right\}\dfrac{M_{\nu^c_2}}{M_{\nu^c_2}-M_{\nu^c_1}}~.
\end{equation} 

It is worth emphasizing  that if the first term dominates, fine tuning is required due to the dependence of the mass splitting in the right-handed neutrino sector. Despite the fact that thermal low scale leptogenesis in most cases requires a tiny mass gap in the heavy states, the second term (first diagram in figure 3),  could accommodate a less constrained mass gap through the suppression due to the existence of Yukawa couplings $\lambda, \kappa^{\prime}$ \cite{Pilaftsis:2003gt,Kang:2006sn,SungCheon:2007nw}. However, due to the heavy Higgs $\bar{H}$ mass included in the loop, this contribution is expected to be suppressed. Simplifying the contributions of the two terms in the above equation, the results are summarized to:

\begin{eqnarray}
\mathit{i)}&&|\epsilon_{1}|\sim \dfrac{M_{\nu^c_2}}{16\pi\langle m_{\nu_D} \rangle^2} \sqrt{\Delta m^2_{\nu_{31}}}\dfrac{M_{\nu^c_2}}{M_{\nu^c_2}-M_{\nu^c_1}}\\
\mathit{ii)}&&|\epsilon_{1}|\sim \dfrac{M_{\nu^c_2}}{16\pi\langle m_{\nu_D} \rangle^2} \sqrt{\Delta m^2_{\nu_{31}}}\dfrac{M_{\nu^c_2}}{M_{\nu^c_2}-M_{\nu^c_1}}\times |\lambda_{ij}^{\nu}|^2|\kappa^{\prime}|^2~.\label{Leptogen}
\end{eqnarray}

\noindent 
These couplings are referring not to the first generation, since the lightest of the sterile neutrino's coupling is bounded by the thermodynamic condition $\Gamma(\nu^c_1)<H(T=M_{\nu^c_1}),$ where H stands for the Hubble expansion. The novelty of the F-theory implementation of the leptogenesis scenario is that fine tuning is not a problem, since the singlets can acquire  appropriate VEVs regulating this way the scale of the produced asymmetry, without the requirement of  $\Delta m_{\nu^c_{21}}\rightarrow0$. The coupling $\kappa^{\prime}$ is suppressed by the string scale, an effect which is absent in the standard field theory GUT framework. 

\section{Neutrinoless double beta decay }

We have already observed in the analysis of the neutrino mass matrix  
 the involvement of new neutral states $s$ which act as  sterile neutrinos.
 Furthermore, the Majorana nature of neutrino states implies violation of lepton number by two units $\Delta L=2$.  The presence of these  ingredients could potentially provide  low energy signals which are worth investigating. Amongst those implications, neutrinoless double beta decay (for a review see~\cite{Vergados:2012xy}) seems a suitable experimental process, where the presence of additional sterile neutrinos could enhance the decay's amplitude and shed some light on the mixing between the active and sterile  sectors. Clearly, within the context of the inverse seesaw mechanism of the present model, the described scenarios of leptogenesis, unitarity violation and double beta decay are entangled  and the goal of this section is to extract some bounds for the mass splitting of the right-handed neutrinos and their Majorana phases.\par
As can be inferred  even a simple extension of the SM with a Majorana mass term  could predict the occurrence of the  $\beta\beta$-decay process through a Lagrangian term of the form 

\begin{align}
	\mathcal{L}\supset \sum_{i=1}^3 g_{F}^2 U_{e_i}^2\gamma_{\mu} P_{R} \dfrac{\slashed{p}+m_i}{p^2-m_i^2}\gamma_{\nu}P_L,
\end{align}
where the $m_i$ represent the masses of the neutrinos and $p$ is the momentum of the virtual particle in the decaying process \footnote{As a matter of fact, this propagator is related to the Nuclear Matrix Element (NME), which is being used to capture the nucleus dynamics - see for example  eq. (3) in \cite{Faessler:2014kka}.}. 

The neutrinoless double beta decay, ${0\nu}\beta\beta$, in the presence of the light neutrinos is described by the effective mass:

\begin{equation}
	m_{ee}=|\sum_{i=1}^3U_{ei}^2m_i|
\end{equation}

\noindent 
In this model, the summation in the above formula is modified in order to accommodate the extended neutrino sector \cite{Abada:2018qok}:

\begin{equation}
	m_{ee}=\sum_{i=1}^{3+n}U_{e_j}^2p^2 \dfrac{m_i}{p^2-m_i^2},\label{nbound}
\end{equation} 
where $U_{e_j}^2$ stands for the mixing of the electron neutrino with the other states and the decay width is proportional to $\Gamma_{0\nu2\beta}\sim m_{ee}$. Recent experimental constraints put a stringent bound on the allowed region \cite{Deppisch:2012nb, DellOro:2016tmg,Faessler:2014kka}, which is:

\begin{equation}
	|m_{ee}|\in  [10^{-3},10^{-1}]\;{\rm eV}~. 
\end{equation}


It is obvious that for high scale masses of the right-handed neutrinos ($m_{\nu^c}\gg  {\rm TeV}$) and intermediate scale sterile singlets ($m_{s}\sim {\rm keV}$), sizable effects on the $0\nu\beta\beta$ decay could be attributed to the mass of heavy neutrinos and the mixing of the various sectors. From \eqref{nbound}, there exist two important limits concerning the mass of the extra neutrinos~\cite{Blennow:2010th, Abada:2018qok}, where the propagator is modified as:

\begin{align}
	\mathit{i)}\; m_i\ll p^2: \dfrac{1}{p^2-m_i^2}=\dfrac{1}{p^2}+&\dfrac{m_i^2}{p^4}+\mathcal{O}(\dfrac{m_i^4}{p^6})~,\label{cond1}\\
	m_{ee}=\sum_{i=1}^{3+n}U_{e_i}^2m_i~,&\\
	\mathit{ii)}\; m_i\gg p^2: \dfrac{1}{p^2-m_i^2}=-\dfrac{1}{m_i^2}&+\mathcal{O}(\dfrac{m_i^4}{p^6})~,\label{cond2}\\
	m_{ee}=-\sum_{i=1}^{3+n}U_{e_i}^2m_i\dfrac{p^2}{m_i^2}~.&
\end{align}

 We are going to analyze the neutrinoless double beta decay in both of these limits. The case \textit{ii}, in particular, represents the seesaw mechanism presented above, but the ``light''  neutrinos (case \textit{i} ) could also be interesting for experiments searching low energy sterile neutrinos. 
  In order to get an insight for the neutrinos sector and reach some representative conclusion, we adopt a tangible strategy and  work in a simplified effective scenario. Thus, for the light neutrinos, it would be reasonable to consider a single neutrino (e.g. the electron neutrino), whilst for the heavier sector we will  assume a case of three neutrinos (two right-handed ones and one sterile). Similar approach has been considered in previous literature ( for a few representative papers, see for example relatable examples with 3+1 or 3+2 neutrinos  in~\cite{Hagedorn:2021ldq,Bolton:2019pcu,Deppisch:2015qwa,Abada:2018qok,Mitra:2011qr}). In~\cite{Bolton:2019pcu}, a similar model was considered, however  the present analysis  considers three different scales (${\rm eV}$-${\rm keV}$-${\rm TeV}$) and as stated above it would be ideal to derive a bound for the mass splitting of the heavy neutrinos, since this fraction is used in leptogenesis. In addition, we are going to sketch the production mechanism of the sterile neutrinos, if they were to be identified as a dark matter particle, through their coupling with the right-handed neutrinos. Consequently, the mixing matrix would be $4\times4$, which can be parameterized as follows:

\begin{align}
	U(\nu_e,\nu_1^c,\nu_2^c,s)=&\left(
	\begin{array}{cccc}
		1 & 0 & 0 & 0 \\
		0 & c_{12} & s_{12} & 0 \\
		0 & -s_{12} & c_{12} & 0 \\
		0 & 0 & 0 & 1 \\
	\end{array}
	\right)
	\left(
	\begin{array}{cccc}
		c_{\text{e2}} & 0 & e^{-i \delta } s_{\text{e2}} & 0 \\
		0 & 1 & 0 & 0 \\
		-e^{i \delta } s_{\text{e2}} & 0 & c_{\text{e2}} & 0 \\
		0 & 0 & 0 & 1 \\
	\end{array}
	\right)
	\left(
	\begin{array}{cccc}
		c_{\text{e1}} & s_{\text{e1}} & 0 & 0 \\
		-s_{\text{e1}} & c_{\text{e1}} & 0 & 0 \\
		0 & 0 & 1 & 0 \\
		0 & 0 & 0 & 1 \\
	\end{array}
	\right)\notag \\
	&\left(
	\begin{array}{cccc}
		c_{\text{es}} & 0 & 0 & s_{\text{es}} \\
		0 & 1 & 0 & 0 \\
		0 & 0 & 1 & 0 \\
		-s_{\text{es}} & 0 & 0 & c_{\text{es}} \\
	\end{array}
	\right)
	\left(
	\begin{array}{cccc}
		1 & 0 & 0 & 0 \\
		0 & c_{\text{s1}} & 0 & s_{\text{s1}} \\
		0 & 0 & 1 & 0 \\
		0 & -s_{\text{s1}} & 0 & c_{\text{s1}} \\
	\end{array}
	\right)
	\left(
	\begin{array}{cccc}
		1 & 0 & 0 & 0 \\
		0 & 1 & 0 & 0 \\
		0 & 0 & c_{\text{s2}} & s_{\text{s2}} \\
		0 & 0 & -s_{\text{s2}} & c_{\text{s2}} \\
	\end{array}
	\right)\cdot \mathcal{\varPhi},
\end{align}
where the last matrix represents the Majorana phases $\varPhi={\rm diag}(1,e^{i \varphi _1},e^{i \varphi _2}, e^{i \varphi _s})$, where $\;\phi\in (0,\pi)$ and $\delta$ is the Dirac phase (this will not play a crucial role, since we treat light neutrinos as a single state) and $s_{ij},c_{ij},\; (i,j=e,1,2,s),\; \theta\in(0,\frac{\pi}{2})$ are the mixing angles between the neutrinos. Now, denoting with 
$\hat{M}(\hat{m}_{\nu},\hat{m}_{\nu_i^c},\hat{m}_s)$ the diagonalized neutrino mass matrix  the following equation holds:

\begin{equation}
	U\hat{M}(\hat{m}_{\nu},\hat{m}_{\nu_i^c},\hat{m}_s)U^T=\cal {M}_{\nu}~.
\end{equation}
where,

\be 
{\cal M}_{\nu}=
\left(\begin{array}{cccc}
	0&m_{\nu_D}&0&0\\
m_{\nu_D}& M_{11} &M_{12}&M_{1s}
\\
0& M_{21} &M_{22}&M_{2s}
	\\
0&M_{1s}&M_{2s}&M_s
\end{array}\right)~.\label{MNreduced}
\ee 
where $M_{ij}$ denote the elements of the  $2\times 2$  right-handed neutrino matrix $M_{\nu_i^c} $ in this example.

Comparing particular elements of the mass matrix $\cal {M}_{\nu}$  with the mass eigenbasis matrix  $\hat{M}(\hat{m}_{\nu},\hat{m}_{\nu_i^c},\hat{m}_s)$ we can extract some useful bounds. First of all, a few assumptions need to be taken into account in order to simplify the calculations. Hence, we will assume that the mixing angles between the active neutrinos $\nu_e$ and the sterile ones $\nu_{1,2}^c, \nu_s$ are small, plus that the masses of the heavy states are much heavier compared to the light and the sterile states: 

\begin{align}
	\theta_{e1},\theta_{e2},\theta_{es}\ll 1\Rightarrow &\;cos(\theta)\cong 1, \;sin(\theta)\cong \theta ,\notag\\
	\dfrac{\hat{m}_{\nu}}{\hat{m}_{1,2}}&,\dfrac{\hat{m}_s}{\hat{m}_{1,2}}\ll 1~.\label{assumption}
\end{align}

Under these assumptions, the sines $(s_{e1},s_{e2},s_{es})$ represent small angles, but we are not going to change their symbols in the calculations below. 
Observing the structure of the neutrino mass matrix $\cal {M}_{\nu}$ given in~(\ref{MNreduced}), we compare  the two zero entries  $\{11\}$,$\{13\}$ and the $\{33\}$ element  $M_s\rightarrow \mu$ with the corresponding ones of $\hat{M}(\hat{m}_{\nu},\hat{m}_{\nu_i^c},\hat{m}_s)$.  These yield the following equations

\begin{align}
	\mathcal{M}_{\nu}^{11}=(U\hat{M}(\hat{m}_{\nu},\hat{m}_{\nu_i^c},\hat{m}_s)U^T)_{11}&=0\label{equ1},\\
	\mathcal {M}_{\nu}^{13}=(U\hat{M}(\hat{m}_{\nu},\hat{m}_{\nu_i^c},\hat{m}_s)U^T)_{13}&=0\label{equ2},\\
	\mathcal {M}_{\nu}^{33}=(U\hat{M}(\hat{m}_{\nu},\hat{m}_{\nu_i^c},\hat{m}_s)U^T)_{33}&=\mu~.\label{equ3}
\end{align}
For (\ref{equ1}) we obtain:
\begin{align}
	\mathcal{M}_{\nu}^{11}=\dfrac{\hat{m}_{\nu}}{\hat{m}_1}e^{-i(\delta+2\phi_2)}-2e^{-i\delta}c_{s1}s_{es}&\big[(e^{i2\Delta\phi_{21}}+c_{s2}^2z-\dfrac{\hat{m}_2}{\hat{m}_1})s_{e1}s_{s1}+e^{-i\delta}zc_{s2}s_{e2}s_{s2}\big]=0,\notag\\
\end{align} 
where  we have introduced the definitions 
\[ 	z=\dfrac{\hat{m}_2}{\hat{m}_1}-\dfrac{\hat{m}_s}{\hat{m}_1}e^{i2\Delta\phi_{21}}\cong \dfrac{\hat{m}_2}{\hat{m}_1} ;\;\; {\rm and} \;\; \Delta \phi_{21}=\phi_2-\phi_1~.\]
Then,
\begin{align}
	\dfrac{s_{e1}}{s_{e2}}=-e^{-i\delta}&\dfrac{\hat{m}_2c_{s2}s_{s2}}{s_{s1}(\hat{m}_1e^{i2\Delta\phi_{21}}-\hat{m}_2s_{s2}^2)}\; ~.\label{11elem}
\end{align}

\noindent 
Since we have assumed only a single light neutrino, the Dirac phase from this point on is taken $\delta= 0$. In this limit,  for small active-sterile angles, we expect the fraction between them to be positive, which can be translated using the denominator of \eqref{11elem} to:

\begin{equation}
	s_{s2}^2>\dfrac{\hat{m}_1}{\hat{m}_2}\cos(2\Delta\phi_{21})~.\label{bound1_0}
\end{equation}

\noindent 
It is readily seen, that, the mixing between the left and right-handed neutrinos are fully determined by the ``dark" sector i.e. the right-handed neutrinos and the sterile singlet. Proceeding to the $\{33\}$ element, a similar analysis leads to the following bounds:

\begin{align}
	\mathcal {M}_{\nu}^{33}=e^{i2\phi_1}\hat{m}_1s_{s1}^2&+c_{s1}^2\big[e^{i2\Delta\phi_{s1}}c_{s2}^2m_s+e^{i2\phi_2}\hat{m}_2s_{s2}^2\big]=\mu,\notag\\
	\dfrac{\mu}{\hat{m}_1}e^{-i2\phi_1}=s_{s1}^2&+c_{s1}^2\big[e^{i2\Delta\phi_{s1}}c_{s2}^2\dfrac{\hat{m}_s}{\hat{m}_1}+e^{i2\phi_2}\dfrac{\hat{m}_2}{\hat{m}_{1}}s_{s2}^2\big]~.\label{33elem}
\end{align}

\noindent 
Now, implementing  the Cauchy-Schwarz theorem 
for  the $\{33\}$ element we obtain:

\begin{align}
	\dfrac{\mu}{\hat{m}_1}\leq s_{s1}^2+c_{s1}^2\big(\dfrac{\hat{m}\hat{m}_s^2}{\hat{m}_1^2}&c_{s2}^4+\dfrac{\hat{m}_2^2}{\hat{m}_1^2}s_{s2}^4+\dfrac{\hat{m}_s\hat{m}_2}{\hat{m}_1^2}\sin(2\theta_{s2})\cos(2\Delta\phi_{s2})\big)^{1/2}\Rightarrow \notag \\
	&c_{s1}^2\leq\dfrac{\hat{m}_1-\mu}{\hat{m}_1-\hat{m}_2s_{s2}^2}, \; s_{s2}^2<\dfrac{\hat{m}_1}{\hat{m}_2},\label{cs1}
\end{align}
where the last inequality has been derived under the assumptions that $\hat{m}_1>\mu$ and $c_{s1}^2>0$. Remarkably, using \eqref{bound1_0}, a very narrow bound  can be derived:

\begin{equation}
	\dfrac{\hat{m}_1}{\hat{m}_2}\cos(2\Delta\phi_{21})<s_{s2}^2<\dfrac{\hat{m}_1}{\hat{m}_2}~.\label{bound}
\end{equation}

\noindent  The inequality~\eqref{cs1} which describes the mixing of the sterile sector, can be written equivalently as:
\begin{equation}
	c_{s1}^2\leq \dfrac{ \frac{\hat{m}_1}{\hat{m}_2}-\frac{\mu}{\hat{m}_2}}{ \frac{\hat{m}_1}{\hat{m}_2}-s^2_{s2}}~. \label{MajPh}
\end{equation}

\noindent Proceeding as previously the equality \eqref{11elem}  yields:
\begin{align}
	\dfrac{\hat{m}_1}{\hat{m}_2}\geq s_{s2}\big(1-\dfrac{s_{e2}c_{s2}}{s_{s1}s_{e1}}\big)~.
\end{align}

\noindent Regarding the Majorana phases from the \eqref{33elem}, the imaginary part of the equation implies:

\begin{equation}
	\dfrac{\sin(2\phi_1)}{\sin(2\Delta\phi_{21})}=-\dfrac{\hat{m}_2}{\mu}c_{s1}^2s_{s2}^2,\label{MajPh1}
\end{equation}

\noindent where this equation is valid only for specific regions for $\phi\in(0,\pi)$.

In figure~4,  we plot the left hand side of equation \eqref{MajPh1}. In the lower right square the two heavy neutrinos have the same (negative) CP charge and represent Majorana fermions. In the upper left square, the heave neutrinos have opposite CP charge and they can form a pseudo-Dirac pair. Considering the case, where the mass scale $\mu \rightarrow 0$, we expect that  lepton number violation is absent and $\Delta L=2$ processes are suppressed.

\begin{figure}[H]

		\centering\includegraphics[scale=0.5]{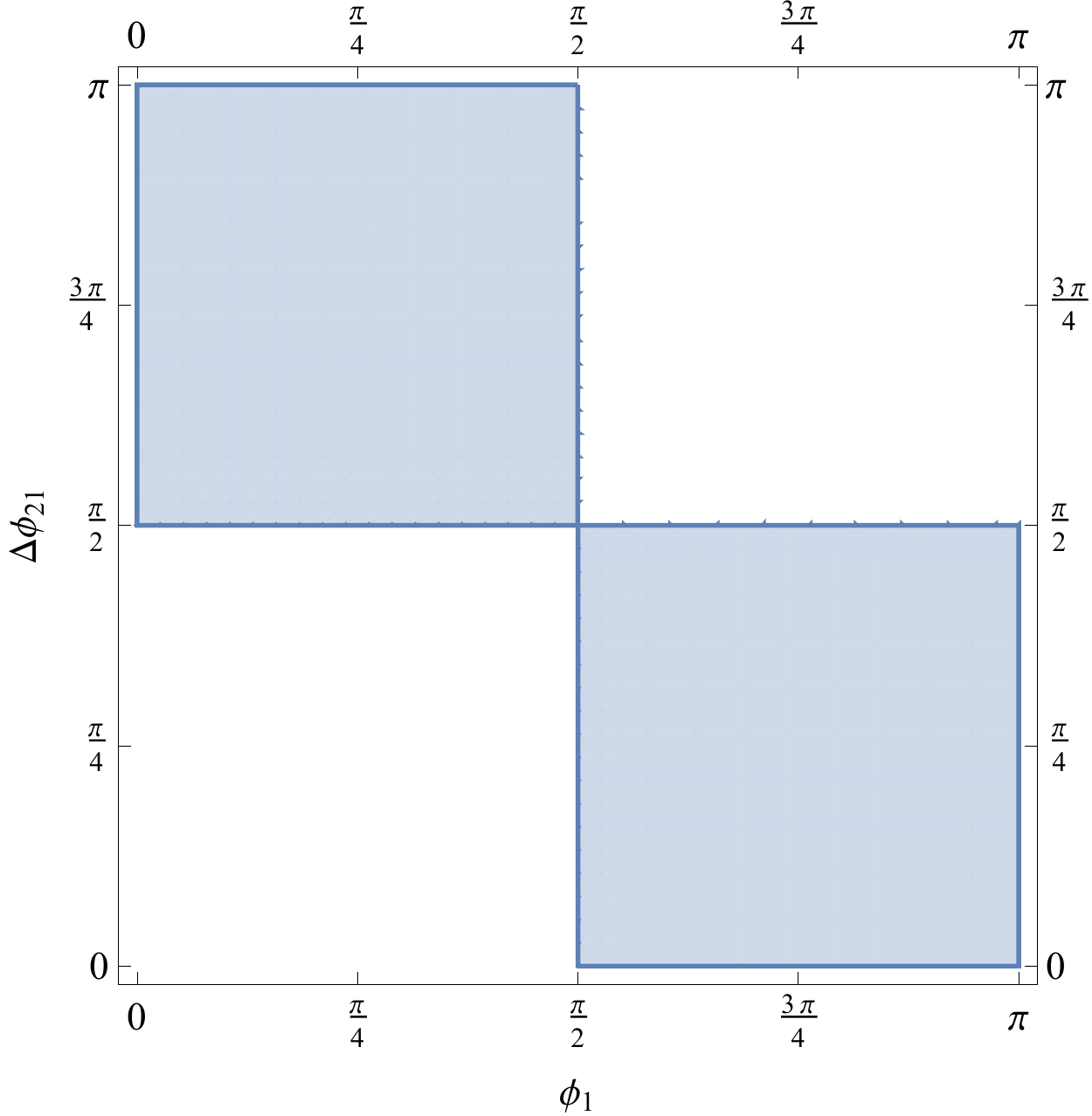}
	
	\caption{The left hand side of the equation \eqref{MajPh1} where we see that that the right-handed neutrinos can have opposite CP charge (upper left square) or the same (lower right square), which would yield interesting phenomenological implications. See  main text.}\label{CP-charge}
\end{figure}

The third and last constraint  to be imposed is associated with  the $\{13\}$ element. This can be used to  constrain 
the mixing $s_{es}$ between the active neutrino 
and the singlet $s$. Thus,  $	\mathcal {M}_{\nu}^{13}=0$ yields 

\begin{align}
\dfrac{s_{es}}{s_{e2}}=\dfrac{s_{e1}}{s_{e2}}s_{s1}c_{s1}\dfrac{-\Delta\hat{m}_{21}+\hat{m}_1e^{i4\Delta\phi_{21}}-\hat{m}_2e^{i2\Delta\phi_{21}}}{\hat{m}_1-c_{s1}^2(\hat{m}_1-\hat{m}_2s_{s2}^2e^{i2\Delta\phi_{21}})}+\mathcal{O}(\dfrac{\hat{m}_{\nu,s}}{\hat{m}_{1,2}}),\label{13elem}
\end{align}
where $\Delta\hat{m}_{21}=\hat{m}_2-\hat{m}_1$, while  for a controllable calculation we have neglected terms suppressed by the heavy neutrinos. After the parametrization of the different mixing angles and the phases, we are in a position to estimate  their impact on the neutrinoless double beta decay. Following the discussion around equations~(\ref{cond1},\ref{cond2}), two distinct regimes can be defined:

\begin{eqnarray}
	\mathit{i)}&&\; m_{ee}= \hat{m}_{\nu_L}+U_{e1}^2\hat{m}_1+U_{e2}^2\hat{m}_2+U_{es}^2\hat{m}_s,\;\hat{m}_i\ll p^2\notag\\
	&&m_{ee}=U_{e2}^2\big(\dfrac{\hat{m}_{\nu_L}}{U_{e2}^2}+\dfrac{U_{e1}^2}{U_{e2}^2}{\hat{m}_1}+{\hat{m}_2}+\dfrac{U_{se}^2}{U_{e2}^2}\hat{m}_s\big),\notag\\
	\mathit{ii)}&&\; m_{ee}= \hat{m}_{\nu_L}-U_{e1}^2\dfrac{p^2}{\hat{m}_1}-U_{e2}^2\dfrac{p^2}{\hat{m}_2}+U_{se}^2\hat{m}_s,\;\hat{m}_i\gg p^2 \notag\\
&&m_{ee}=U_{e2}^2\big(\dfrac{\hat{m}_{\nu_L}}{U_{e2}^2}-\dfrac{U_{e1}^2}{U_{e2}^2}\dfrac{p^2}{{\hat{m}_1}}-\dfrac{p^2}{{\hat{m}_2}}+\dfrac{U_{es}^2}{U_{e2}^2}\hat{m}_s\big),
	\label{NDB}
\end{eqnarray} 

\noindent 
where in both regimes the amplitude is defined up to an overall factor, but the terms in the parentheses are in principle responsible for the process. The mixing matrices $U_{ei}^2$ for small angles can be represented by the sines ($U_{ei}^2\rightarrow s_{ei}$) computed before, so from the previous analysis we know every fraction (see equations (\ref{11elem},\ref{13elem}) appearing  in the formulas. We have neglected the mixing of the left handed neutrinos, since we have used only the electron neutrino. Consequently, the whole process is parametrized up to an overall factor $U_{e2}^2$. It is worth noticing that $\frac{U_{es}^2}{U_{e2}^2}=\gamma \frac{U_{e1}^2}{U^2_{e2}}$,

\begin{equation}
	\gamma=s_{s1}c_{s1}\dfrac{-\Delta\hat{m}_{21}+\hat{m}_1e^{i4\Delta\phi_{21}}-\hat{m}_2e^{i2\Delta\phi_{21}}}{\hat{m}_1-c_{s1}^2(\hat{m}_1-\hat{m}_2s_{s2}^2e^{i2\Delta\phi_{21}})},
\end{equation}
 simplifying both of the parentheses in equation \eqref{NDB} as:

\begin{align}
	\mathit{i)}&\; m_{ee}=U_{e2}^2\big(\dfrac{\hat{m}_{\nu_L}}{U_{e2}^2}+\hat{m}_2+\dfrac{U_{e1}^2}{U_{e2}^2}(\hat{m}_1+\gamma\hat{m}_s)\big)>0\notag\\
	\mathit{ii)}&\; m_{ee}=U_{e2}^2\big(\dfrac{\hat{m}_{\nu_L}}{U_{e2}^2}-\dfrac{p^2}{\hat{m}_2}+\dfrac{U_{e1}^2}{U_{e2}^2}(-\dfrac{p^2}{\hat{m}_1}+\gamma\hat{m}_s)\big)>0
\end{align}

The requirement of having positive mass for the $m_{ee}$ leads the quantities in the parentheses to be bounded as:

\begin{align}
	\mathit{i)}&\dfrac{\hat{m}_{\nu_L}}{U_{e2}^2}+\hat{m}_2>-\dfrac{U_{e1}^2}{U_{e2}^2}(\hat{m}_1+\gamma\hat{m}_s)\Rightarrow \gamma<-\dfrac{\hat{m}_1}{\hat{m}_s}\notag\\
	\mathit{ii)}&\dfrac{\hat{m}_{\nu_L}}{U_{e2}^2}-\dfrac{p^2}{\hat{m}_2}>\dfrac{U_{e1}^2}{U_{e2}^2}(\dfrac{p^2}{\hat{m}_1}-\gamma\hat{m}_s)\Rightarrow \gamma>\dfrac{p^2}{\hat{m}_1\hat{m}_s}~.\label{gamma}
\end{align}

Since we expect a positive fraction \eqref{13elem} for the mixing angles, we must also have $\gamma>0$.  Hence the first case above is incompatible,  since the assumptions stated in \eqref{assumption} imply $\gamma<0$. In the second case a bound for the $\gamma$ variable is extracted, which  is going to be used to define the allowed parametric region for the neutrinoless double beta decay. In order to get an insight for the leptogenesis scenario regarding the nature of right-handed neutrinos participating in it, we need to check the asymptotic region of the fraction $\frac{\hat{m}_1}{\hat{m}_2}\rightarrow (0,1)$. In the vanishing  mass limit, the $\frac{s_{e1}}{s_{e2}}\gamma$ variable reduces to:

\begin{equation}
\frac{s_{e1}}{s_{e2}}\gamma=-2\dfrac{\cos ^2\left(\Delta \phi _{21}\right)}{\cos(2\Delta \phi_{21})} \dfrac{c_{s2}}{c_{s1}s^3_{s2}} \Rightarrow \Delta \phi_{21}\in (\frac{\pi}{4},\frac{\pi}{2}) \cup (\frac{\pi}{2},\frac{3\pi}{4})~.
\end{equation}

In this limit, neutrinoless double beta decay scans the Majorana nature of the right-handed neutrinos and if baryon asymmetry is explained through leptogenesis, it is expected to happen due to the lightest heavy neutrino as in equation \eqref{leptcase1}. Inversely stated, if two sterile neutrinos are observed, the mass fraction and their relative CP-charge difference can be used in order to extract the scale of neutrinoless double beta decay and the scale of possible sterile singlet through the analysis above.

In the degenerate mass limit  $\frac{\hat{m}_1}{\hat{m}_2}\rightarrow 1$, some useful conclusions can be extracted with respect to the mixing of the sterile neutrinos with the two heavy states. In this case the $\frac{s_{e1}}{s_{e2}}\gamma$ variable is written as

\begin{equation}
\frac{s_{e1}}{s_{e2}}\gamma=\frac{c_{\text{s1}} c_{\text{s2}} \left(\cos \left(2 \Delta \phi _{21}\right)-\cos \left(4 \Delta \phi _{21}\right)\right) s_{\text{s2}}}{\left(\cos \left(2 \Delta \phi
   _{21}\right)-s_{\text{s2}}^2\right) \left(c_{\text{s1}}^2 \left(\cos \left(2 \Delta \phi _{21}\right) s_{\text{s2}}^2-1\right)+1\right)}~.
\end{equation}

As it can be observed in the numerator above, there is a sign flip in the region of $\Delta \phi_{21} \in(\frac{\pi}{3},\frac{2\pi}{3})$, where in this region the sterile singlet couples stronger with the second sterile neutrino $\theta_{s1}>\theta_{s2}$. Hence, in this limit if the two sterile neutrinos are observed with $\Delta \phi_{21} \in (0,\frac{\pi}{2})$, the neutrinoless double beta decay is expected to be suppressed due to the Pseudo-Dirac pair, while in the  $\Delta \phi_{21} \in (\frac{\pi}{2},\pi)$ they represent two Majorana fermions with degenerate mass.

\par We are going to present the masses of the neutrinos for the singlet VEVs, whose values are shown in Table~\eqref{Vevs}. For these particular VEVs, the neutrinos are computed through the case $\beta)$ \eqref{TwoCases} of section 6., the leptogenesis through the case $ii)$ \eqref{Leptogen} and the neutrinoless double beta decay is expected at the degenerate mass limit (Table \ref{Table 6}).

\begin{table}[H]
	\begin{center}  \small%
		\begin{tabular}{|p{1.7cm}|p{1.7cm}|p{1.7cm}|p{1.7cm}|p{1.7cm}| p{1.7cm}|p{1.7cm}|p{1.7cm}|}
			\hline
			$\hat{m}_{\nu_i}\;({\rm eV})$& $\hat{m}_{\nu^c}\;({\rm GeV})$  & $\hat{m}_s\;({\rm keV})$ & $\epsilon_1$  & $\eta$ & $\theta_{\nu s}$ \\
			\hline
			$0.1$ & $4.3\times 10^{14}$ & $0.55$ & $2.3\times 10^{-6}$ & $2.1\times 10^{-3}$ & $4.7\times 10^{-4}$ \\
			\hline
		\end{tabular}
	\end{center}
	\caption{Masses computed for the following scales: $ m_{\nu_D} =174\;{\rm GeV}$, $M_{\nu^c}=4.3\times 10^{14} \;{\rm GeV}$, $M_s=19.1\;{\rm keV}$, $M_{\nu^cs}=89.3\times10^3 \;{\rm GeV}$, $\Delta m_{31}^2=2.2\times 10^{-3} \;{\rm eV^2}$, and the first and second generation of heavy neutrinos at $\big(1.8\times 10^{10},3\times 10^{10}\big)\; {\rm GeV}$. Regarding the neutrinoless double beta decay, the model probes the blue region of $\frac{\hat{m}_1}{\hat{m}_2}\rightarrow 0.6$.}\label{Table 6}
\end{table} 

\begin{figure}[H]
    \centering
    \begin{minipage}{.5\textwidth}
        \centering
        \includegraphics[scale=0.75]{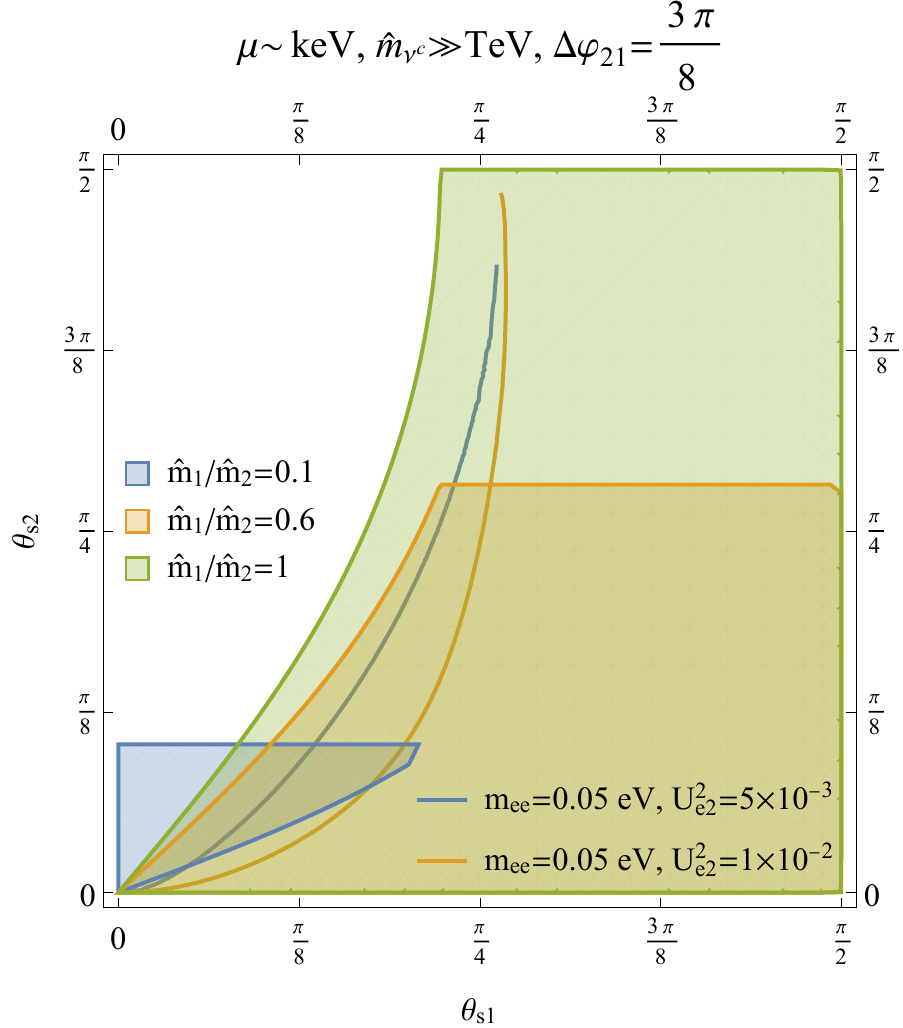}
    \end{minipage}%
    \begin{minipage}{0.6\textwidth}
        \centering
        \includegraphics[scale=0.75]{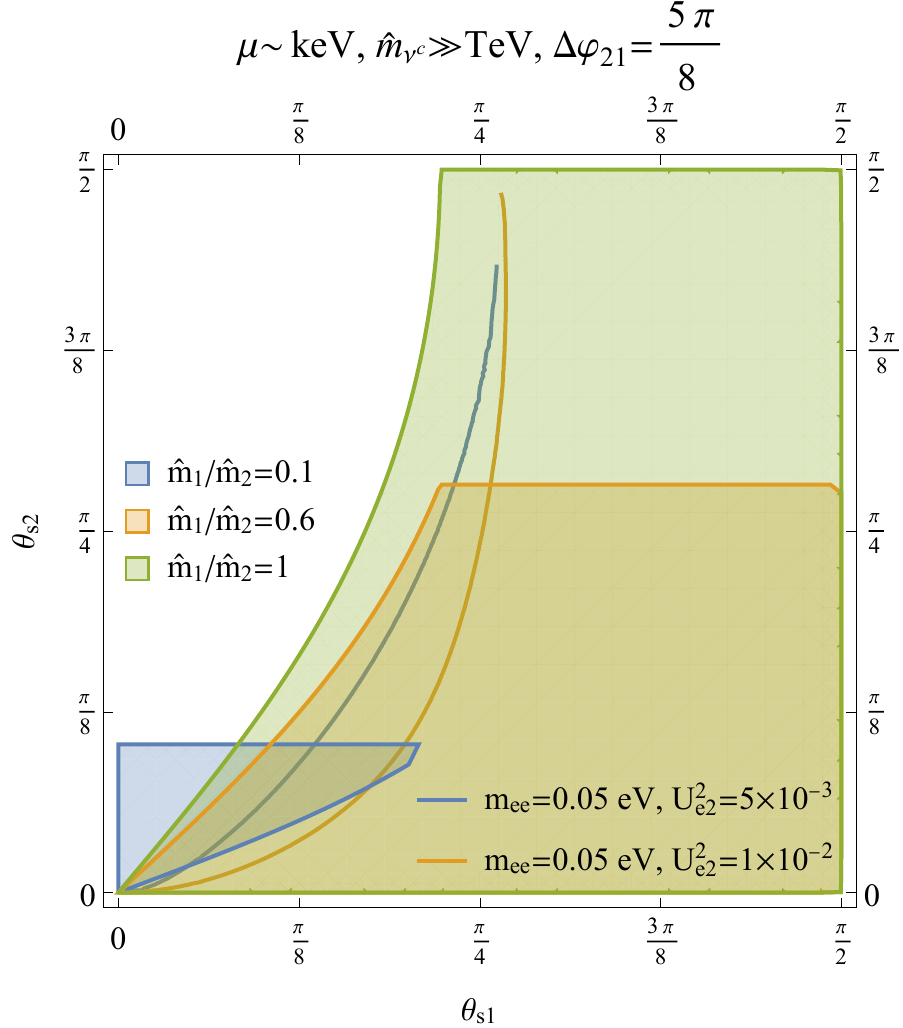}
    \end{minipage}
    \caption{The shaded region depicts the allowed parameter space defined by the inequalities \eqref{bound},\eqref{MajPh},\eqref{gamma} and the curves represent the solutions for the neutrinoless double beta decay from the equation \eqref{NDB}.}  \label{Fig5}
\end{figure}

Also, in the two plots of figure~\ref{Fig5}  a couple of solutions of the equation \eqref{NDB} are depicted for various values of $U^2_{e2}$ and the effective electron neutrino mass $m_{ee}$. 

\section{On the muon magnetic moment $g_{\mu}-2$}

The extra vector-like states appearing in the zero-mode spectrum of the F-theory flipped $SU(5)$ are a possible
source of the $g_{\mu}-2$ enhancement~\cite{Li:2021cte,Ellis:2021vpp}. The relevant couplings are 

\begin{equation}
\mathcal{W}=\lambda^{\prime}\bar{h}h \dfrac{\langle\chi \psi^2\rangle}{M_{S}^3}\bar{\psi}+ \lambda_{ij}^e e^c_i \bar f_j h+ \alpha_{mj} \bar E^c_m e^c_j\,\bar \psi +\beta_{mn} \bar E^c_mE^c_n \,\bar \zeta+\gamma_{nj}E^c_n\bar f_j h\chi~.
\end{equation}
which give rise to the one-loop graph shown in figure~6.

\begin{figure}[H]
\centering
\begin{tikzpicture}
\begin{feynman}

\vertex (a){\(l_i\)};
\vertex[right=of a](b);
\vertex[above right=of b](c);
\vertex[below right=of c](d);
\vertex[right=of d](e)  {\( e^c\)};
\vertex[above right=0.5 cm and 0.4 cm of c](f)  {\(\langle \psi^2 \rangle\)};
\vertex[above left=0.5 cm and 0.4 cm of c](g)  {\(\langle \chi \rangle\)};
\vertex[above=1.0cm of c](w) {\(\langle H_u
 \rangle\)};

\diagram*{
	(a)--[fermion](b),
	(b)--[scalar, bend left,edge label=\(H_d\)](c),
	(b)--[majorana, insertion=0.5, edge label'=\(M_{\bar{E}^cE^c}\)](d),
	(c)--[scalar, bend left, edge label=\(\bar{\psi}\)](d),
	(e)--[fermion](d),
	(c)--[scalar](f),
	(c)--[scalar](g),
	(c)--[scalar](w)
};

\end{feynman}
\end{tikzpicture}
\caption{Feynman diagram for the contribution of the vector-like pair in the $g_{\mu}-2$ process}
\end{figure}
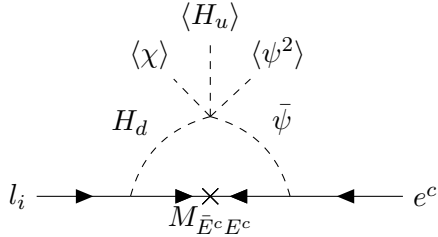

   Its contribution to $g_{\mu}-2$ is highly dependent on the mass of the additional vector-like lepton-type charged  singlets $E^c,\bar E ^c$,  since the latter participate in the loop.
   In the  model under consideration their mass is given in terms of the VEV of the singlet $\bar{\zeta}$, i.e.,  $M_{\bar{E}^cE}=\langle\bar{\zeta}^2\rangle$. It is also worth mentioning that, the very same VEV appears in the proton decay process, where the masses of the Higgs triplets are assigned a high scale mass due to this singlet. Consequently, low scale supersymmetry could not be a viable choice, in the case we would like to have a substantial contribution to $\Delta \alpha_{\mu}\sim \frac{m_{\mu}\langle h\rangle}{\langle \bar{\zeta}^2\rangle}$. Split susy fits better in such a scenario, where the mass of vector-like singlets can be lowered down to TeV scale and sufficiently explain the $g_{\mu}-2$ discrepancy. Although, due to the mixing of the vector like leptons with the leptonic sector of the model, a mass matrix is constructed as it is shown in Table \ref{Table 7}.

\begin{table}[H]
\begin{center} \small%
\begin{tabular}{p{1cm}|p{3.5cm} p{3cm}}
$M^2_{Ee}$ & $E^c$ & $e^c_j$\\
\hline
$L_i$ & $\dfrac{\langle h \rangle \chi}{M_{str}}$ & $\langle h \rangle$\\ \hline
$\overline{E}^c$ &$\bar{\zeta}$ & $\bar{\psi}$
\end{tabular}
\caption{Mixing between the vector like leptons and the electrons.}\label{Table 7}
\end{center}
\end{table}

In this case, the resulting mass of the states, which contribute in the above process could in principle be around ${\rm TeV}$ scale.

\begin{align}
m_{1}=\dfrac{\langle h\rangle\chi}{M_{str}}\cos^2(\theta)-\dfrac{\langle h \rangle+\bar{\zeta}}{2}\sin(2\theta)+\bar{\psi}\sin^2(\theta)\notag\\
m_{2}=\bar{\psi}\cos^2(\theta)+\dfrac{\langle h \rangle+\bar{\zeta}}{2}\sin(2\theta)+\dfrac{\langle h\rangle\chi}{M_{str}}\sin^2(\theta)~.
\end{align}

For the singlet VEVs mentioned at the previous sections, there are in principle light states after the mixing between the electrons and the vector-like singlets. Consequently, the heaviest of these singlets will lay at $\rm{TeV}$ scale, contributing to the $g_{\mu}-2$ sufficiently to explain the discrepancy. Using the vevs of the model described before, the contribution to the $g-2$ anomaly can be summarized to the following calculation as:

\begin{equation}
\Delta \alpha_{\mu}\sim \dfrac{m_{\mu}\langle h\rangle}{m_{2}^2}\sim \dfrac{105 \times 10^{-3} \;174 \; \rm{GeV^2}}{(89.3\times 10^3)^2 \rm{GeV^2}}\sim 23 \times 10^{-10}
\end{equation}

 \section{A possible interpretation of the CDF measurement of the W-mass}

 Recently,  the CDF II collaboration~\cite{CDF:2022hxs} using data collected in proton-antiproton collisions  at the Fermilab Tevatron collider, has measured  the W-boson mass to be  $m_{W}= 80,433.5\pm  9.4\, {\rm MeV}/c^2$. This value is in glaring discrepancy with the SM prediction, and the LEP-Tevatron combination which is 
 $M_W = 80,385 \pm  15\, {\rm MeV}/c^2$.   Since then several SM and MSSM extensions with the inclusion of new particles have been proposed to explain theoretically the experimental prediction of the W-mass.
 Taking the CDF result at face value,   in the following we will show how the new ingredients in the present flipped $SU(5)$ construction
 may predict this W-mass enhancement. We first 
 recall that the neutrino mass matrix  formed by the three left- and right-handed  neutrinos,  as well as the sterile ones, is diagonalized by a unitary transformation. However, the mixing matrix diagonalizing the effective $3\times 3$ light neutrino mass matrix  obtained after the implementation of the
 inverse seesaw mechanism, need not be unitary.  Consequently, this can in principle lead to a  non-unitary leptonic mixing matrix
 which in section 6 has been parametrized as $V_{\ell} = (1+\eta ) U_{PMNS}$.  We will see that such effects can in principle modify the mass of the W-boson.
 
 In the context of the Standard Model, the mass of the W-boson can be inferred by comparing the muon decay prediction with
 the Fermi model~\cite{Heinemeyer:2004gx}
 \be 
 M_W^2 \left(1-\frac{M_W^2}{M_Z^2}\right)= \frac{\pi \alpha_{em}}{\sqrt{2} G_F} (1+\Delta r)~,\label{FermiSMmatch}
 \ee
 where $\alpha_{em}$ and $ G_F$  are  the  fine structure and Fermi  constants respectively, and $ \Delta r$  stands for all possible  radiative corrections~\cite{Sakurai:2022hwh,Lopez-Val:2014jva}. Once $\Delta r$ is known, the  SM prediction of the W-boson mass is obtained by solving the formula~(\ref{FermiSMmatch}).  However, in the present case 
 the non-unitarity in the PMNS matrix affects drastically the muon decays and consequently the measurement of the muon lifetime.
 The precise knowledge of these effects are  
 essential  since they determine the Fermi constant $G_F$ 
 which is involved  in the determination of the $W$ and $Z$  boson masses. Thus,  one  might  expect possible  deviations from the $G_F$ value   when measured ($G_{\mu}$)  in muon decay.  The non-unitary corrections are connecting them according to ~\cite{Fernandez-Martinez:2016lgt,Blennow:2022yfm}:
 
 \begin{equation}
 	G_{F}=G_{\mu}(1+\eta_{ee}+\eta_{\mu \mu }),\label{GFmodif}
 \end{equation} 
 
 \noindent 
 where $\eta_{ee},\eta_{\mu \mu}$ are the $\{11\},\{22\}$ elements of the unitarity violation matrix $\eta$. 
 Implementing the above formula for the Fermi constant, and solving (\ref{FermiSMmatch}), the mass of the W-boson is given by
 
 \begin{equation}
 	M_W^2= \frac{1}{2}\left(M_Z^2+\sqrt{1-\frac{4\pi \alpha_{em}(1-\eta_{\mu \mu }-\eta_{ee})}{\sqrt{2}G_{\mu}M_Z^2}(1+\Delta r)}\right)~,
 \end{equation}

 \noindent 
 Clearly, a possible increment of the W-mass may arise either due to non-unitarity inducing  positive $\eta_{ee,\mu\mu}$ contributions, or from possible suppression of the radiative corrections  $\Delta r$.
 
 Notice that $\Delta r$ can also receive additional corrections 
 due to  the pair $E^c+\bar E^c$ appearing in the flipped $SU(5)$ spectrum. 
 Their couplings in the superpotential induce a Wilson coefficient 
 $(C_{h\ell})_{ij} =-\lambda_i\lambda_j^*/(4 m_E^2)$  which gives  a sufficient contribution to the $W$-mass  for $M_E\sim 5$ GeV~\cite{Bagnaschi:2022whn,Endo:2022kiw}

Using the bounds for the mixing angles and the $\eta_{\alpha\beta}$ elements  from Table IV of~\cite{Fernandez-Martinez:2016lgt}, we can plot the mass of the W-boson in terms of the non-unitary effects, where it  is clearly seen that for small deviations from the unitary form of the leptonic mixing matrix can explain the experimental result. From the diagonalization of the neutrino matrix \eqref{NMatrix1}, we expect two forms for the unitarity violation, corresponding to the two cases mentioned there. These two cases are

\begin{align}
\alpha)\;\; \eta\cong \mathcal{O}(\dfrac{1}{2}\dfrac{m^2_{\nu_D}}{M^2_{\nu^cs}}),\qquad \beta)\;\; \eta\cong  \mathcal{O}(\dfrac{1}{2}\dfrac{m^2_{\nu_D}(M^2_{\nu^cs}+M_s^2)}{(M^2_{\nu^cs}-M_{\nu^c}M_s)^2})~.\label{eta2}
\end{align}

Since we are interested in the second case, it is obvious that the scale $M_s$, which is responsible for the lepton number violation will play a crucial role. The specific form (texture) of the fermion mass matrices, of course, can in principle produce different -model dependent- scenarios of the unitarity violation. Despite this, we can derive the scale of the $\eta$ matrix and extract some preliminary insights for the experimental signal. In figure~7, we plot the mass of the W-boson for different values of the lepton number violating scale $M_s$. As it is pointed out in~\cite{Blennow:2022yfm},  the insertion of right-handed neutrinos in the model produces a positive definite $\eta$ matrix  which  is a necessary condition to explain the CDF-measurement of the W-boson mass. In fact a small lepton number violation can accommodate the W-mass discrepancy. Notably, at the same time, the sterile states can explain the Cabibbo angle anomaly~\cite{Coutinho:2019aiy} through the mixing term $ \kappa_i\overline{H} F_i s \,\bar\psi$, although, the Cabibbo angle anomaly is not completely related to neutrinos, but to the inert singlet states involved in the seesaw mechanism.

\begin{figure}[H]
	
	\centering\includegraphics[scale=0.75]{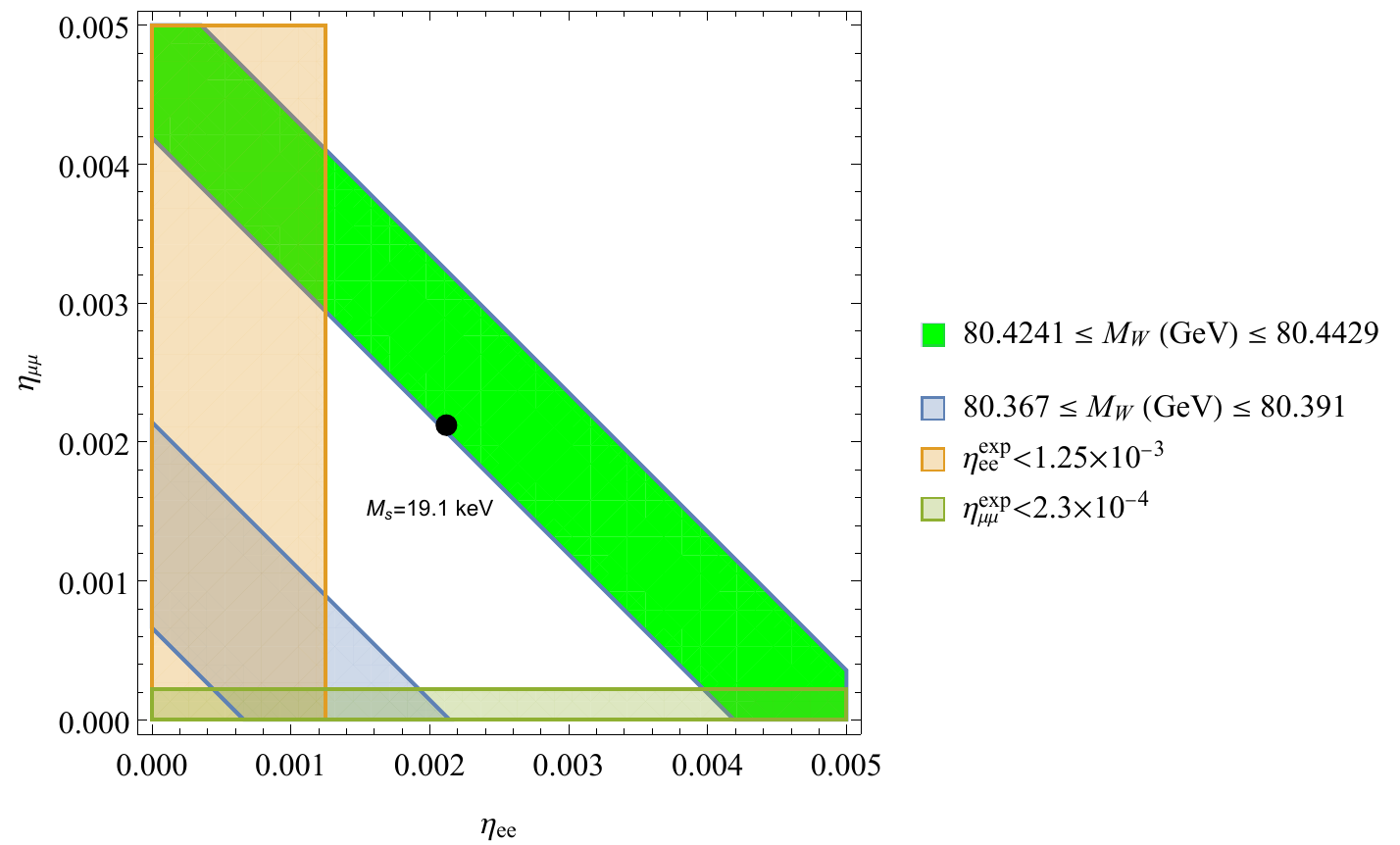}
	
	\caption{ Plot of case $\beta)$ $\eta$ \eqref{eta2} (black dots under the assumption $\eta_{ee}\sim \eta_{\mu\mu}$), using $ m_{\nu_D} =174\;{\rm GeV}$, $M_{\nu^c}=4.3\times 10^{14} \;{\rm GeV}$, $M_s=19.1\;{\rm keV}$, $M_{\nu^cs}=89.3\times10^3 \;{\rm GeV}$. Blue shaded region is the previous W-boson mass and green is the current measurement.}
\end{figure}

It is readily seen from the above that unitarity violation plays a crucial role in the mass of the W-boson. The main characteristic of the inverse seesaw mechanism~\footnote{We note that another solution with Type III seesaw with the presence of an $SU(2)$ Higgs triplet has been also suggested~\cite{Ghoshal:2022vzo}.}  is the small violation in the lepton number by the scale $M_s$. Large deviations from the PMNS-matrix can occur in the case where the sterile neutrinos lay at an intermediate scale (${\rm keV}-{\rm MeV}$),  since there is significant mixing between those states with the active neutrinos. In conclusion, one could conjecture  that the neutrino masses, or more specifically the violation in the lepton number, play a significant role in the LFV physics, where sterile states allow this type of processes to evade the GIM suppression of SM. In conclusion, under the above mentioned circumstances, the rich structure of the F-theory flipped $SU(5)$ may suggest a viable interpretation of the W-mass increment~\footnote{In the context of F-theory, a different explanation with D3 branes has been suggested in~\cite{Heckman:2022the}.}. \par

\begin{figure}[H]
    \centering
    \begin{minipage}{.5\textwidth}
        \centering
        \includegraphics[scale=0.7]{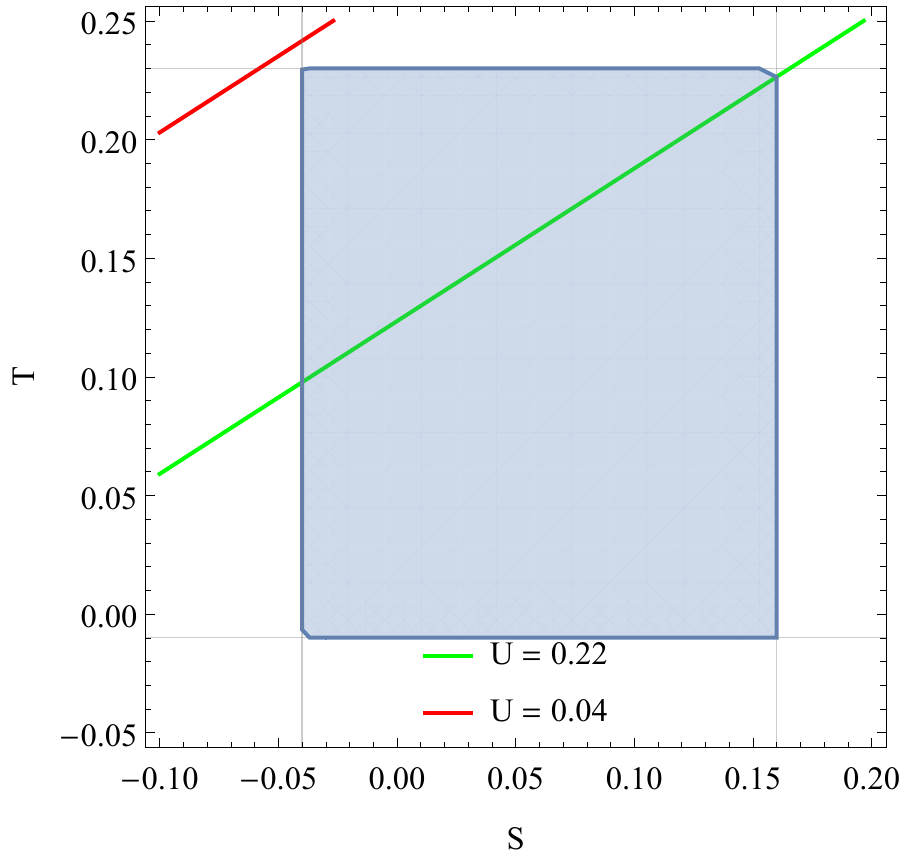}
    \end{minipage}%
    \begin{minipage}{0.6\textwidth}
        \centering
        \includegraphics[scale=0.7]{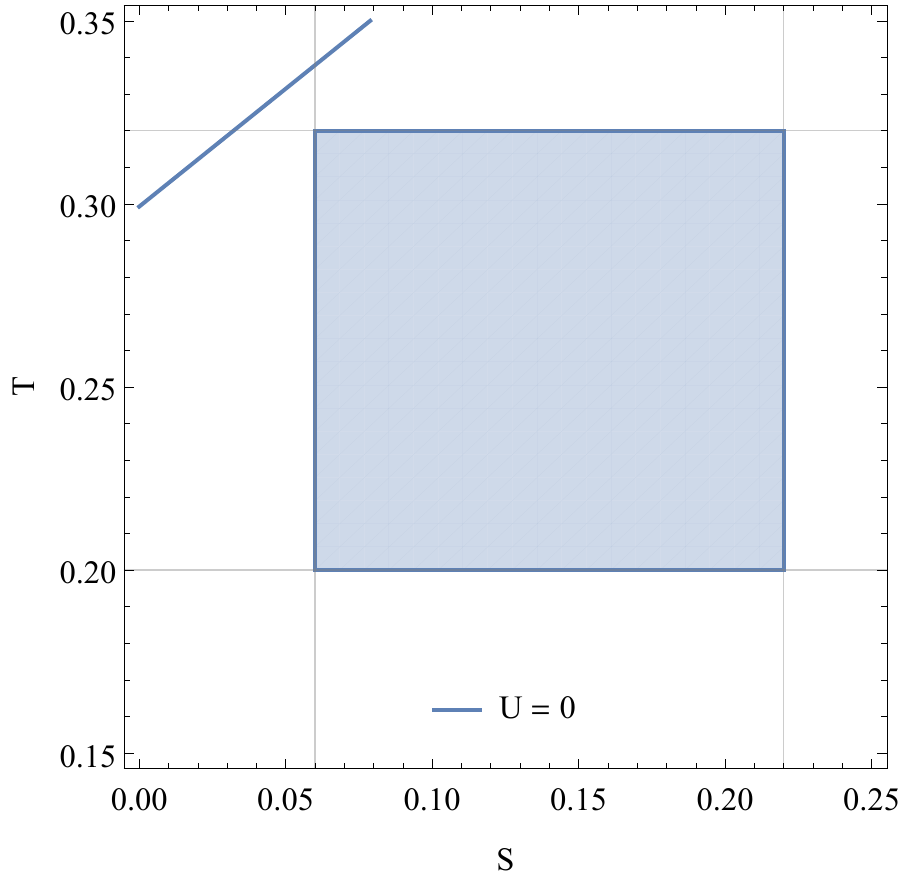}
    \end{minipage}
    \caption{Left: Solution for S,T parameters with fixed parameter U, where the blue shaded region covers the bounds, as obtained by fit taking into account the new mass of W boson. Right: No solutions found when U is vanishing.} \label{Figure 8}
\end{figure}

As for the oblique parameters, which parameterize the effects of new physics in the electroweak observables, they have a direct implication on the recently observed mass shift of the W boson. Following the work of \cite{Ciuchini:2013pca} with respect to the mass of W boson and \cite{Lu:2022bgw} for the recently obtained fit on the oblique parameters, we could test our model and the unitary violation as a proposed solution.

\begin{align}
\frac{M_W^{new}}{M_W}=-\frac{a \left(-\frac{U \left(c_W^2-s_W^2\right)}{2 s_W^2}-2 c_W^2 T+S\right)}{4 \left(c_W^2-s_W^2\right)}-\frac{\Delta G s_W^2}{2 \left(c_W^2-s_W^2\right)}+1,\label{Obliq}
\end{align}

\noindent 
where $s_W^2=1-\frac{M_W^2}{M_Z^2}$ and the $\Delta G$ is the modification of the Fermi constant $G_F=G_{\mu}(1+\Delta G)$. So, in our scenario, $\Delta G$ can be identified with the unitarity violation terms $\Delta G= \eta_{ee}+\eta_{\mu \mu}$. In the two figures below, we plot equation \eqref{Obliq} for various values of the $S,T$ parameters with a fixed $U$.  So, after inserting $\Delta G=2\times 2.1\times 10^{-3}$ and the masses of the W, Z bosons, the solutions are depicted below (Fig. \ref{Figure 8}).

\begin{align}
&S\in (-0.04,0.16),\;T\in(-0.01,0.23),\; U\in (0.04,0.22)\notag \\
&S\in (0.06,0.22),\;T\in(0.2,0.32), U=0
\end{align}

\section{Gauge coupling unification and Yukawa couplings}

For the RGE's analysis of our model, we consider a low energy spectrum of the MSSM model accompanied by the presence of the vector-like singlets $E^c$. Starting with the beta function concerning the MSSM and the flipped $SU(5)$ particle content (for beta functions of flipped  see for example~\cite{Ellis:1988tx,Leontaris:1992cf}), we summarize the formulas below:

\begin{align}
b_1&=\frac{3}{5} \left(\frac{3 n}{10}+\frac{1}{2}n_H\right)+n_v\notag \\
b_2&=-6 + 2 n + \frac{1}{2}n_H+n_v\notag \\
b_3&=-9 + 2 n+n_v,\notag \\
b_5&=\frac{3 n_{10}}{2}+\frac{n_5}{2}+2 n-15\notag\\
b_{1_{\chi}}&=\frac{n_{10}}{4}+\frac{n_5}{2}+2 n
\end{align}
where $n$ is the number of generations and $n_v$ is the number of vector-like families. We can easily  deduce that for $n=3, n_v=0$ we get the usual beta functions of the MSSM:

\begin{align}
\{b_1,b_2,b_3\}=\{\dfrac{33}{5},1,-3\}
\end{align}

After inserting a vector-like pair in the low energy spectrum, we can plot the running of the coupling constants at 1-loop level and we can, eventually, spot the unification point. After the insertion of the parameter $a=\frac{g^2}{4\pi}$, we get

\begin{align}
a_i^{-1}(Q)=a_i^{-1}(Q_0)-\dfrac{b_i}{2\pi}\log(\dfrac{Q}{Q_0}),
\end{align}
where the effect of a vector-like singlet family in the model in the beta functions is $\Delta b_i^{MSSM}=\{1,1,1\}$. There are two energy regions: from $0<\mu<M_Z$, we run the beta functions of the SM, from $M_Z<\mu<M_{E^c}$ we run the MSSM plus the vector like particles and finally we run the flipped $SU(5)$ till a unification point.   Plotting the running parameters of the model, we can see in the following plot that the unification scale is about $M_{GUT}\sim 10^{17}\;\rm{GeV}$.

\begin{figure}[H]  
\centering\includegraphics[scale=0.7]{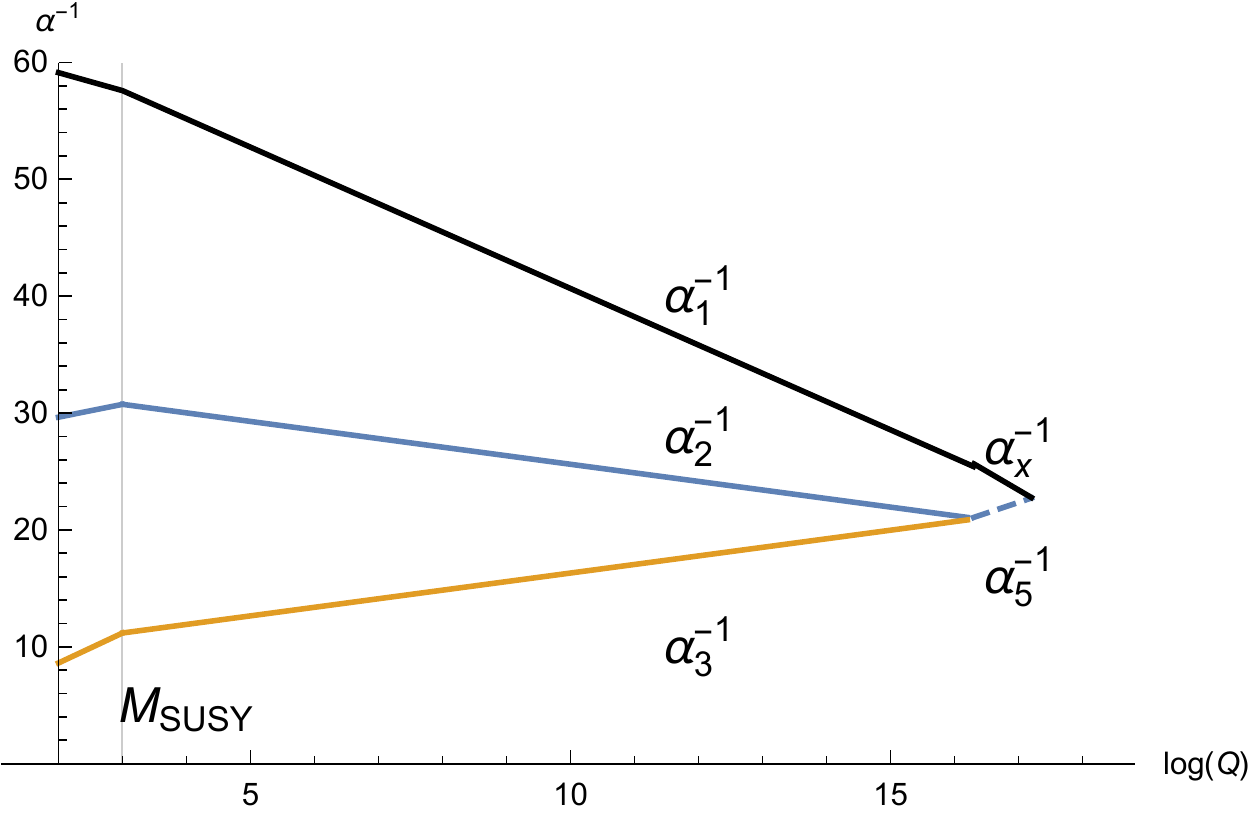}
\end{figure}

The unification scale is at $M_{U}\cong  10^{17} \rm{GeV}$, where the couplings constants are 

\begin{align}
\alpha_1^{-1}(M_Z)=59.38,\;\alpha_2^{-1}(M_Z)=29.74,\;\alpha_3^{-1}(M_Z)=8.44,\; \alpha^{-1}_U=22.5
\end{align}

\begin{figure}[H]
\centering\includegraphics[scale=1]{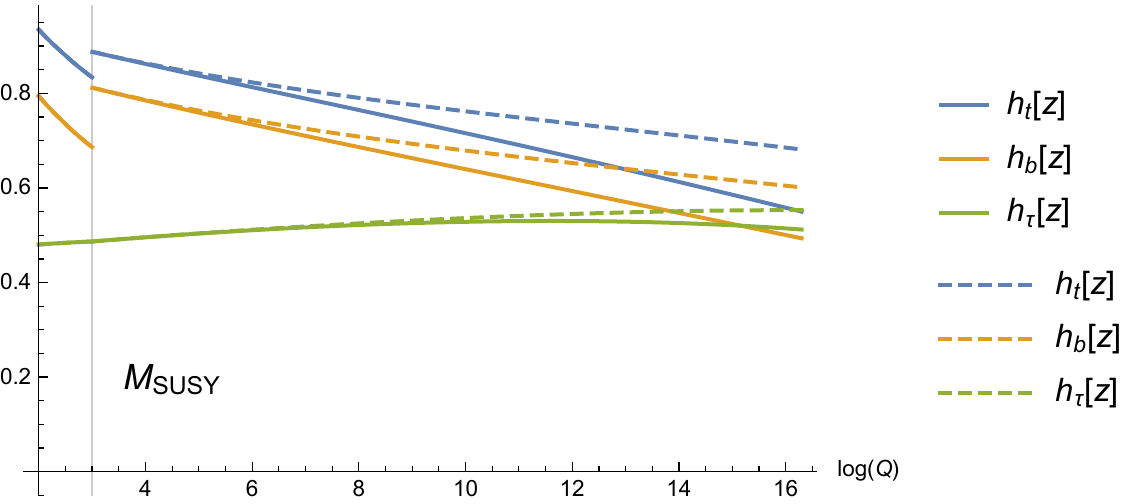}
	
\caption{Yukawa evolution for the following parameters SUSY parameters $m_g=2 \;\rm{TeV},\mu=0.5\;\rm{TeV},\tan\beta=58,m_t=3\;\rm{TeV},h_t(0)=0.94,h_b(0)=0.8,h_{\tau}(0)=0.48.$ The dashed lines are the Yukawa without the vector like families where they deviate for $\tan\beta>50$ as expected. The thick lines present the Yukawa couplings evolution with the insertion of a vector like family.}\label{Figure 9}
\end{figure}

As for the Yukawa couplings, we only consider the third generation (where the for the top, bottom quarks and the $\tau$ lepton are denoted as $h_{t},h_{b},h_{\tau}$ respectively) and the mixing effects of the abelian $U(1)$ symmetries ,during the evolution down to the low energy values, are being neglected. For the computation, the Mathematica code SARAH-4.15.0 \cite{Staub:2013tta} was used and the following plot depicts with thick lines the running of the spectrum with the vector-like family, where the dashed line contains the same information without the additional particles. During the computation, we have taken into account that the largest correction due to loops of sparticles is affecting the bottom Yukawa coupling as:
\begin{align}
\delta h_b\cong \dfrac{g_3^2}{12\pi^2}\dfrac{\mu m_g \tan\beta}{m_b^2}+\dfrac{h_t^2}{32\pi^2}\dfrac{\mu A_t \tan\beta}{m_t^2},
\end{align}
where $m_b=\frac{m_{b_1}+m_{b_2}}{2},\;m_t=\frac{m_{t_1}+m_{t_2}}{2}$ are the average masses of the top and bottom squark. Consequently, we could safely extract the conclusion that even at high energies, Yukawa couplings stay under control at a perturbative regime (Fig. \ref{Figure 9}).

\section{Conclusions}

There  is accumulating evidence that the Standard Model spectrum and its minimal supersymmetric extensions require a substantial and 
radical overhaul to account for New Physics phenomena predicted in major experimental facilities around the globe.  Grand Unified 
Theories emerging from String Theory suggest a robust framework where such issues can be addressed by virtue  of new 
ingredients appearing at the effective theory level in a well-defined and consistent way. 
In this work we have constructed an effective low energy model with $SU(3)\times SU(2)\times U(1)$ gauge symmetry derived from 
an $SO(10)$  geometric singularity of an elliptically fibred  CY fourfold over a threefold base.

The first stage of symmetry breaking of the corresponding $SO(10)$ gauge group is realized with an abelian flux along the $U(1)_{\chi}$ factor inside  $SO(10)$, giving rise to the flipped $SU(5)\times U(1)_{\chi}$ model.
At the second stage, this symmetry breaks down to the SM gauge group when a $10_{-2}+\ov{10}_{2}$ pair of $SU(5)\times U(1)$ Higgs multiplets develop VEVs.  As in the standard field theory flipped model~\cite{Antoniadis:1987dx},
the down type colour triplets of these Higgs representations pair up with the triplets in $5+\bar 5$ Higgs multiplets, and receive large masses so that  dimension-five baryon violating operators are adequately suppressed.  
Furthermore, there are several phenomenological predictions associated with extra matter fields which are present in the effective model. Thus, in addition to the MSSM fields, the low energy spectrum contains an extra pair of right-handed  singlets with electric charges $\pm 1$  which contribute to $g_{\mu}-2$. Moreover, extra neutral singlet fields acquire Yukawa couplings with the right-handed neutrinos realizing  an inverse seesaw mechanism.  Taking advantage  of the  parameter space, left unconstrained by flatness conditions and other stringy restrictions,  we assume various limiting cases and single out those ones supporting a viable leptogenesis scenario. 
We further discuss the double beta decay process and pay particular attention to contributions stemming from the mixing effects of the active neutrinos with the  inert singlet fields. 
We illustrate the main points by performing a detailed analysis in a scenario with three active neutrinos ($\nu_e,\nu^c_1,\nu^c_2$)
and one sterile neutral singlet field, and derive constraints on the mixing effects among them. 
We find parametric regions with substantial contributions to $O\nu\beta\beta$ decay rate which could be observed in future experiments.  Finally we discuss deviations from unitarity of the 
effective $3\times 3$ lepton mixing matrix $U_{PMSN}$  and their possible implications on the recently observed deviation of the W-boson mass by CDF II collaboration.

  \section{Appendix}

Consistency with supersymmetry and anomaly cancellation  requires 
that the singlet VEVs  are subject to F- and D-flatness conditions.
The following hierarchy of scales is assumed $\langle H\rangle 
\sim \langle \ov{H}\rangle \sim M_{GUT} \cong  M_{str}$. The singlet VEVs
are also assumed to be smaller than the string scale $ M_{str}$.

 Using the identification~(\ref{singlets}) and $Z_2$ monodromy, the  Yukawa lagrangian for the singlet fields is
 \ba 
 {\cal W}_{\cal S}=
  \lambda_{1} \bar\chi\bar\zeta\psi +\lambda_{2} \bar\psi\zeta\chi  +M_{s}s^2+M_{\chi}\bar\chi\chi+M_{\psi}\bar\psi\psi + M_{\zeta}\bar\zeta \zeta ~.
 \ea 
 The mass scales $M_{\zeta}, M_{\chi}$ etc are assumed to be arbitrary and will be fixed through the flatness conditions. The F-flatness equations
 are
 
 \begin{align}
 &\dfrac{\partial W_{\mathcal{S}}}{\partial \chi}=0\Rightarrow\lambda_2 \bar{\psi}\zeta+M_{\chi}\bar{\chi}=0\notag\\
 &\dfrac{\partial W_{\mathcal{S}}}{\partial \psi}=0\Rightarrow\lambda_1 \bar{\chi}\bar{\zeta}+M_{\psi}\bar{\psi}=0\notag\\
 &\dfrac{\partial W_{\mathcal{S}}}{\partial \zeta}=0\Rightarrow\lambda_2 \bar{\psi}\chi+M_{\zeta}\bar{\zeta}=0\notag\\
 &\dfrac{\partial W_{\mathcal{S}}}{\partial \bar{\chi}}=0\Rightarrow\lambda_1 \bar{\zeta}\psi+M_{\chi}\chi=0\notag\\
 &\dfrac{\partial W_{\mathcal{S}}}{\partial \bar{\psi}}=0\Rightarrow\lambda_2 \chi \zeta+M_{\psi}\psi=0\notag\\
 &\dfrac{\partial W_{\mathcal{S}}}{\partial \bar{\zeta}}=0\Rightarrow\lambda_1 \bar{\chi}\psi+M_{\zeta}\zeta=0~.
 \end{align}

 \noindent The  D-term flatness constraint needs, also, to be imposed which has the following form:
 
 \begin{align}
& \sum_{i\neq j} q_i(\theta_{ij}^2-\theta_{ji}^2)=-c M_{str}^2\Rightarrow \notag\\
 q_{\chi}(\chi^2-\bar{\chi}^2)&+q_{\psi}(\psi^2-\bar{\psi}^2)+q_{\zeta}(\zeta^2-\bar{\zeta}^2)=-c M_{str}^2~.
 \end{align}

In order to derive a solution to the flatness condition, we need to impose  the following conditions

\begin{equation}
M_{\chi}=-\lambda_1 M_{\psi},\;q_i=1~.
\end{equation}
Then, we obtain
\begin{align}
\chi&=\dfrac{M_{\zeta}\rho}{\lambda_1\lambda_2 \sigma},\;\; \bar{\chi}=\dfrac{M_{\psi}\sigma}{\rho}\notag\\
\psi&=-\dfrac{M_{\zeta}}{\lambda_1},\;\; \bar{\psi}=\dfrac{M_{\psi}\lambda_1}{\lambda_2}\notag\\
\zeta&=\dfrac{M_{\psi}\sigma}{\rho},\;\; \bar{\zeta}=-\dfrac{M_{\psi}\rho}{\sigma}\notag \\
\rho=\big((M_{\zeta}^2+cM_{str}^2\lambda_1^2)&\lambda_2^2-\lambda_1^4M_{\psi}^2\big)^{1/2},\;\;\sigma=\big(\lambda_1^2M_{\psi}^2-M_{\zeta}^2\big)^{1/2}~.
\end{align}

Demanding the $\mu$-term ($\chi$ singlet) and $\bar{\psi}$ to lay at the TeV scale, we are going to derive some bounds on the parameters above.

\begin{align}
\dfrac{\bar{\chi}}{\zeta}=1,\;\dfrac{\chi}{\psi \bar{\zeta}}=\dfrac{1}{M_{\psi}},\; \bar{\psi}=\dfrac{M{_\psi}\lambda_1}{\lambda_2},\;M_{\psi}\gg 1~.
\end{align}

So, the corresponding bounds for the parameters are:

\begin{align}
\dfrac{\lambda_2}{\lambda_1}\ll \dfrac{M_{\psi}}{\bar{\psi}\sim TeV},\; M_{\zeta}^2<M_{\psi}^2\lambda_1^2,\; c>\dfrac{M_{\psi}^2\lambda_1^4-M_{\zeta}^2\lambda_2^2}{\lambda_1^2M_{str}^2}~.
\end{align}
 


\section{Additional Models}

In this paper we have explored a flipped $SU(5)$ model based on a specific 
choice of fluxes and choosing a particular matter curve to accommodate the Higgs fields. However, there are other choices which may lead to somewhat modified 
phenomenological implications. Here we present two possible modifications.

We may change the Higgs doublets  of the model, discussed in the main text by choosing the fluxes $M^1_{10}\rightarrow M^2_{10}=1$, so the new Higgs fields are
\begin{align}
h_{-t_1-t_4},\;\;\bar{h}_{t_1+t_3},
\end{align}
\begin{align}
W_{matter}=&\lambda^{u}_{ij}F_i\bar{f}_j\bar{h}\bar{\psi}+\lambda^d_{ij}F_iF_jh\bar{\psi}+\lambda^e_{ij}e^c_i\bar{f}_j h\bar{\psi}+k_i\bar{H}F_is\bar{\psi}\notag\\
&+a_{mj}\bar{E}^c_me^c_j\bar{\psi}+\beta_{mn}\bar{E}^c_mE^c_n\bar{\zeta}+\gamma_{nj}E^c_n\bar{f}_jh\bar{\zeta},
\end{align}
\begin{align}
W_{higgs}=&\lambda_{\mu}\bar{\zeta}(1+\lambda^{'}_{\mu}\bar{H}H\bar{\zeta})\bar{h}h+\lambda_{\bar{H}}\bar{H}\bar{H}\bar{h}\bar{\psi}\bar{\zeta}+\lambda_{H}HHh(\chi\bar{\zeta}+\bar{\zeta}^2\psi)~.
\end{align}

An alternative model with non-zero flux $P$ is the following:
\begin{table}[H]
	\begin{center}  \small%
		\begin{tabular}{|p{1cm}|p{1.0cm}|p{1.cm}|p{1cm}|p{1.0cm}| p{1.0cm}|p{1cm}|p{1cm}|}
			\hline
			$M_1$& $M_3$  & $M_4$ & $P$& $P_5$ &$P_7$ & $M_{10}^1$ & $M_{10}^2$ \\
			\hline
			$3$ & $-1$ & $1$ & $-1$ & $-2$ &$1$ & $1$ & $-1$\\
			\hline
		\end{tabular}
	\end{center}
	\end{table}
	
	This leads to the matter field assignment:
\begin{align}
&10_{t_1}(F_i):3\times(Q,d^c_i,\nu^c_i),\;\; \bar{5}_{t_1}(\bar{f}):2\times (u^c_i,L_i),\;\;\bar{5}_{t_3}(\bar{f^{'}}):1\times (u^c_3,L_3)\notag\\
&1_{t_1}:4\times(e^c_i),\;\; 1_{t_4}:2\times(E^c),\;\; 1_{-t_3}:-3\times (\bar{E}^c),\;\;5_{-2t_1}:1\times h,\;\;\bar{5}_{t_1+t_4}:1\times \bar{h},
\end{align}
	
	The superpotential for the matter fields is 
\begin{align}
W_{matter}=&\lambda^{u}_{ij}F_i\bar{f}_j\bar{h}\chi+\lambda^{'u}_{ij}F_i\bar{f^{'}_j}\bar{h}+\lambda^d_{ij}F_iF_jh+\lambda^e_{ij}e^c_i\bar{f}_jh+\lambda^{'e}_{ij}e^c_i\bar{f^{'}_j}h\chi+\notag\\&+ k_i\bar{H}F_is\bar{\chi}+a_{mj}\bar{E}^c_me^c_j\bar{\chi}+\beta_{mn}\bar{E}^c_mE^c_n\zeta+\gamma_{nj}E^c_n\bar{f}_jh\psi\notag\\
&+\gamma^{'}_{nj}E^c_n\bar{f^{'}_j}h\chi\psi,
\end{align}
and for the Higgs
\begin{align}
W_{higgs}=\lambda_{\mu}\psi(1+\lambda^{'}_{\mu}\bar{H}H\zeta)\bar{h}h+\lambda_{H}HHh(\psi^2+\chi^2\zeta^2)+\lambda_{\bar{H}}\bar{H}\bar{H}\bar{h}\bar{\chi}\zeta~.
\end{align}

\end{document}